\documentclass[aps,prd,groupedaddress,nofootinbib,
superscriptaddress,preprint,12pt]{revtex4-1}
\pdfoutput=1
\usepackage{color}
\usepackage[table]{xcolor}
\usepackage[colorlinks=true,citecolor=blue,linkcolor=blue,
breaklinks=true]{hyperref}
\usepackage{amsmath,amssymb}
\usepackage{graphicx}
\usepackage{slashed}
\usepackage{url}
\usepackage{enumitem}
\usepackage{multirow}
\usepackage{array}
\usepackage{lipsum}
\usepackage{tabulary}

\newcolumntype{?}{!{\vrule width 1pt}}
\newcolumntype{M}[1]{>{\centering\arraybackslash}m{#1}}
\newcommand{\lag}{\mathcal L}

\def\mET{\slashed{E}_T}

\begin{document}
\preprint{HRI-RECAPP-2021-006}
\title{Search for a light $Z^\prime$ at LHC in a neutrinophilic $U(1)$ model}

\author{Waleed Abdallah}
\email{awaleed@sci.cu.edu.eg} 
\affiliation{Regional Centre for Accelerator-based Particle Physics, 
Harish-Chandra Research Institute, HBNI, \\
Chhatnag Road, Jhunsi, Prayagraj (Allahabad) 211\,019, India}
\affiliation{Department of Mathematics, Faculty of Science, Cairo University, Giza 12613, Egypt}

\author{Anjan Kumar Barik}
\email{anjanbarik@hri.res.in}
\affiliation{Regional Centre for Accelerator-based Particle Physics, 
Harish-Chandra Research Institute, HBNI, \\
Chhatnag Road, Jhunsi, Prayagraj (Allahabad) 211\,019, India}

\author{Santosh Kumar Rai}
\email{skrai@hri.res.in}
\affiliation{Regional Centre for Accelerator-based Particle Physics, 
Harish-Chandra Research Institute, HBNI, \\
Chhatnag Road, Jhunsi, Prayagraj (Allahabad) 211\,019, India}

\author{Tousik Samui}
\email{tousiksamui@hri.res.in}
\affiliation{Regional Centre for Accelerator-based Particle Physics, 
Harish-Chandra Research Institute, HBNI, \\
Chhatnag Road, Jhunsi, Prayagraj (Allahabad) 211\,019, India}

\begin{abstract}
We consider a neutrinophilic $U(1)$ extension of the standard model~(SM) which couples only to SM isosinglet 
neutral fermions, charged under the new group. The neutral fermions couple to the SM matter fields through Yukawa interactions. 
The neutrinos in the model get their masses from a standard inverse-seesaw mechanism while an added scalar sector 
is responsible for the breaking of the gauged $U(1)$ leading to a light neutral gauge boson~($Z'$), which has minimal 
interaction with the SM sector. We study the phenomenology of having such a light $Z'$ in the context of 
neutrinophilic interactions as well as the role of allowing kinetic mixing between the new $U(1)$ group with the 
SM hypercharge group. We show that current experimental searches allow for a very light $Z'$ if it does not 
couple to SM fields directly and highlight the search strategies  at the LHC. We observe that multilepton final states
in the form of $(4\ell + \mET)$ and $(3\ell + 2j + \mET)$ could be crucial in discovering such a neutrinophilic 
gauge boson lying in a mass range of $200$--$500$~GeV.    
\end{abstract}

\maketitle
\section{Introduction}
The modern era of particle physics has seen an extremely successful period with the model 
accounting for three of fundamental interactions of nature via gauge symmetries, i.e., the standard model~(SM) of 
particle physics. The SM successfully explains most phenomena involving the elementary particles in nature which have 
been corroborated through observations in dedicated experiments. The discovery of a 125~GeV 
scalar~\cite{Aad:2012tfa,Chatrchyan:2012ufa} {\it viz.} the Higgs boson has completed the hunt for all particles predicted 
in the SM. Despite the remarkable success of the SM, there still remain several unexplained observations from experiments 
that hint at the possibility of new physics beyond the SM~(BSM). One such anomaly is the observation of nonzero mass 
and mixing of neutrinos from neutrino oscillation 
experiments~\cite{Fukuda:1998mi,Aguilar:2001ty,Ahn:2002up,Abe:2011sj,An:2013zwz}.
The otherwise massless neutral fermion within the SM can, in competing BSM extensions,
have either Dirac or Majorana type mass, which is something yet to be established.  A large number of scenarios 
exist to explain observed neutrino masses and 
mixings~\cite{Minkowski:1977sc,Mohapatra:1979ia,GellMann:1980vs,Yanagida:1979as,10.1007/978-1-4684-7197-7_15}
and these possibilities lead to interesting phenomenology of the resulting neutrino mass models~\cite{GonzalezGarcia:2007ib}. 
Besides the neutrino mass puzzle, another curiosity that intrigues us is the true nature of the scalar that has 
been observed at the Large Hadron Collider~(LHC). The complete confirmation of it being the SM Higgs will only 
be possible, once its 
interactions are precisely measured. Until then it does leave the possibility of new physics within the scalar sector as a 
vital area of interest. There are a vast number of BSM theories including some for neutrino mass models, which include 
an extended scalar sector beyond the SM Higgs doublet. Our focus would be on the type which is central to neutrino 
mass models. 

The minimal extension of the scalar sector is usually done with or without a new gauge group, although an extended 
scalar sector is more natural in extended gauge models where the scalars are charged under the new gauge group and 
are responsible for the spontaneous breaking of the new gauge symmetry. 
All such extensions predict some new phenomena that are to be observed in ongoing and upcoming experiments. 
Extension of the SM with an additional Higgs doublet is one of the most popular extension of the SM and popularly 
known as the two Higgs doublet models~(2HDM). In some models the second Higgs doublet is used to give Dirac masses 
to the light neutrinos by introducing new right-handed neutrinos. Such models are popularly called neutrinophilic 
2HDM~($\nu$2HDM)~\cite{Ma:2000cc,Gabriel:2006ns,Davidson:2009ha}, which lead to interesting 
phenomenology and signatures at 
experiments~\cite{Gabriel:2008es,Davidson:2010sf,Haba:2011nb,Chao:2012pt,Maitra:2014qea,Huitu:2017vye}. 
Another popular extension of the SM is the extension with a new $U(1)$ gauge group. The introduction of new gauge 
groups have a different type of consequence in terms of the signature of the model. One immediate consequence is the 
prediction of a new massive gauge boson~($Z'$) after the symmetry breaking of the new $U(1)$ symmetry. 

We all know that $Z'$ bosons~\cite{Langacker:2008yv} are among the very well motivated new physics 
scenarios in the study of BSM physics. The fact that the all successful SM is a gauge symmetry begs the question 
for the BSM to belong to an extended gauge symmetry with the simplest being the addition of a $U(1)$. 
There are numerous examples of models extending the SM gauge symmetry group by an additional $U(1)$ factor, 
which can arise, for example, from grand unified theories where the group of higher rank is broken down to 
the lower rank SM gauge group, leading to an additional $U(1)$ symmetry arising naturally, or in bottom-up approaches 
where the additional $U(1)$ is added to alleviate problems in models of dynamical symmetry breaking, 
supersymmetry (for example the $\mu$ problem), extra dimensions, flavor physics, etc. and can also act as mediators 
for hidden sectors (for extensive reviews see Refs.~\cite{Leike:1998wr,Rizzo:2006nw,Langacker:2008yv}). 
There have also been proposals for neutrino mass generation, for example in $U(1)_{B-L}$ 
extension~\cite{Marshak:1979fm,Mohapatra:1980qe,Khalil:2006yi}. A discovery of $Z'$ and its decays could therefore 
lead us to an understanding of the underlying gauge charges the particles carry, which could give hints to the 
underlying physics BSM (as the conditions of the new symmetry being anomaly free leads to specific 
charge assignments). However, there is currently no experimental evidence of such a $Z'$, which could have two 
possibilities. $Z'$ may be very heavy to be discovered at current energies and we need to go for higher energies 
in its search, or it may be light but couples very weakly to the SM particles (similar to the SM Higgs search). 
We consider the latter possibility in this work while also invoking the novelty of the model providing a solution to 
the neutrino mass puzzle, leading us to a twofold motivation to consider such an extension. As the LHC has not observed a 
signal for new physics, proposing a light $Z'$ in such extensions is quite difficult unless it weakly couples to the SM sector. 
In this model, which is trivially anomaly free, we can naturally have a light $Z'$ while ensuring a popular seesaw mechanism 
for neutrino mass. We also need not tune the gauge couplings to unnaturally small values for a light $Z'$ unlike for 
example in $U(1)_{B-L}$ models, as this extension allows the gauge couplings to be of similar strength to any SM 
gauge coupling.

We consider an extra $U(1)$ symmetry under which the SM particles are sterile. This is more in the line of a hidden 
extra $U(1)$ considered before in another context by one of us~\cite{Grossmann:2010wm,Das:2015enc}. Only new 
SM isosinglet fermions, an electroweak~(EW) singlet scalar and a neutrinophilic Higgs doublet speak to this extra $U(1)$. 
These new fields act as messenger particles between the $U(1)$ and the SM sector. The extra $U(1)$ symmetry is broken 
at the EW scale by the vacuum expectation value~(VEV) of an EW singlet Higgs boson along with the second Higgs 
doublet. Thus the model predicts a heavy $Z'$ at the EW scale along with additional neutral fermions and 
scalar particles. We show through this work that the prediction of such an extension of the SM which can explain the 
light neutrino mass and with a particle spectrum that has minimal interactions to the charged fermions has its 
own set of challenges of observation and how such a scenario can be observed in the ongoing collider experiments. 

The search for $Z'$ boson has been extensively studied at the LHC where most of the searches put 
strong limits on the mass of the $Z'$ based on its interaction properties~\cite{Langacker:2008yv, Accomando:2013ita}. 
The most popular channel to search for $Z'$ is usually the dilepton channel which gives stringent constraint on the production 
of $Z'$ at the LHC~\cite{Aaboud:2017buh,Sirunyan:2019wqq}. However, in our model, an interesting scenario arises where 
the $Z'$ can be significantly lighter than current limits and can evade bounds from the existing $Z'$ search. 
For such a $Z'$ we find that the multilepton channel proves much more promising. In this study, we mainly focus on $Z'$ 
from the viewpoint of its neutrinophilic nature.\footnote{Similar models in the context of an ultralight mediator 
with cosmological implications and neutrino phenomenology have been studied before~\cite{He:2020zns,Berbig:2020wve}.} 

The paper is organized as follows. In Sec.~\ref{sec:model} we briefly discuss the framework 
of the $U(1)$ gauged neutrinophilic model and calculate the mass and mixing parameters 
for the scalar, gauge and fermion sectors in the model. In Sec.~\ref{sec:constraints} we  
discuss the relevant theoretical and experimental constraints before we move on to 
Sec.~\ref{sec:collider} where we present the LHC analysis of the model in the $4\ell$ and $3\ell$ rich final states
coming from the $Z'$ mediated heavy neutrino production. Finally we summarize and conclude in 
Sec.~\ref{sec:summary}.

\section{The Model} \label{sec:model}

The model is an extension of the SM where the gauge group is augmented with an extra $U(1)_X$ gauge group 
and four new fields, {\it viz.} a second Higgs doublet~($H_2$), a scalar singlet~($S$), and two chiral sterile 
neutrinos~($N_L,\,N_R$) added for each generation.
All the new fields are charged under the gauge group $U(1)_X$ while all the SM particles are neutral. 
The charge assignments of the new particles along with the first Higgs doublet~($H_1$), which is the SM Higgs doublet, 
are listed in Table~\ref{tab:charges}.
\begin{table}[!h]
\begin{center}
\begin{tabular}{|c|c|c|c|c|c|}
\hline 
Fields  & $SU(3)_C$ & $SU(2)_L$ & $U(1)_Y$ & $U(1)_X$ & Spin \\[1mm]
\hline 
$H_1$  & 1 & 2 & $-1/2$ & 0 & 0 \\ [1mm]
\hline 
$H_2$ & 1 & 2 & $-1/2$ & $-\,q_x$ &  0 \\ [1mm]
\hline 
$S$ & 1 & 1 & 0 & $2q_x$ &  0 \\ [1mm]
\hline 
$N_L^i$ & 1 & 1 & 0 & $q_x$  & 1/2 \\ [1mm]
\hline 
$N_R^i$ & 1 & 1 & 0 & $q_x$ & 1/2 \\ [1mm]
\hline
\end{tabular}
\end{center}
\caption{New scalar~($H_a, S$, $a$=1,2) and matter~($N_L^i, N_R^i$, $i$=1,2,3) fields and their charge assignments
under the SM gauge group and $U(1)_X$.}
\label{tab:charges}
\end{table}
Looking at the charge assignments, it is quite clear why we refer the model as a neutrinophilic one.
The new isosinglet charge-neutral fermions are the only spin-$1/2$ fields which carry a $U(1)_X$ charge 
and therefore would lead to couplings of the new gauge boson with the neutrinos after symmetry breaking.

With the assigned charges, the most general gauge invariant Lagrangian that can be added to the SM 
Lagrangian, is given by
\begin{eqnarray}
\lag &\supset& \left(D_\mu H_1 \right)^\dagger D_\mu H_1 + \left(D_\mu H_2 \right)^\dagger D_\mu H_2 + \left(D_\mu S \right)^\dagger D_\mu S - \mu_1 H_1^\dagger H_1 - \mu_2 H_2^\dagger H_2 - \mu_s S^\dagger S \nonumber \\
&       & +\ i\,\overline N_L \gamma^\mu D_\mu N_L + i\,\overline N_R \gamma^\mu D_\mu N_R - 
\hat{M}_N\left(\overline{N}_L N_R + \overline{N}_R N_L \right) -\left\{ Y_\nu\,\overline l_L H_2 N_R + H.c.\right\} \nonumber\\ 
&       & -\ \lambda_1 \left(H_1^\dagger H_1\right)^2 - \lambda_2 \left(H_2^\dagger H_2\right)^2 - \lambda_{12} H_1^\dagger H_1 H_2^\dagger H_2 - \lambda'_{12} \left|H_1^\dagger H_2\right|^2\nonumber \\
&       & -\ \lambda_s \left(S^\dagger S\right)^2 - \lambda_{1s} H_1^\dagger H_1 S^\dagger S - \lambda_{2s} H_2^\dagger H_2 S^\dagger S - \left\{Y_R S \overline N_R N_R^C + Y_L S \overline N_L N_L^C + {\rm\, H.c.}\right\} \nonumber\\
&       & +\left\{\mu_{12} H_1^\dagger H_2 + {\rm\, H.c.} \right\}.
\label{eqn:lag}
\end{eqnarray}

Note that the last term in the Lagrangian breaks the $U(1)_X$ symmetry
explicitly. This soft-breaking term is needed to give mass to the pseudoscalar after the 
symmetry breaking. In addition, the singlet scalar $S$ plays a crucial role in defining the mechanism for 
neutrino mass generation, notwithstanding the fact that it is also responsible for the mass of the $U(1)_X$ 
gauge boson. We shall now discuss the mass and mixings of the scalars, gauge bosons and matter fields 
following the spontaneous symmetry breaking of the gauge symmetries.

\subsection{Masses and mixing of the scalars}
\label{sec:sclars}
The $U(1)_X$ symmetry is spontaneously broken when either the singlet $S$ or the doublet $H_2$ acquires 
a VEV while the SM gauge symmetry breaks when either of the two Higgs doublets 
get a VEV. The Higgs doublets and the scalar singlet fields can be redefined by shifting with their VEVs in the usual 
way. Defining the VEVs for the Higgs doublets and singlet $S$ as $v_1,~v_2,$ and $v_s$, respectively, we can
rewrite the fields as follows: 
\begin{eqnarray}
H_1 = \begin{pmatrix} \dfrac{v_1+\rho_1+i\eta_1}{\sqrt 2} \\ \phi_1^-\end{pmatrix} \, , \qquad
H_2 = \begin{pmatrix} \dfrac{v_2+\rho_2+i\eta_2}{\sqrt 2} \\ \phi_2^-\end{pmatrix} \, , \qquad
S   = \dfrac{v_s+\rho_s+i\eta_s}{\sqrt 2} \, .  \label{eqn:vevs}
\end{eqnarray}

In order for the potential to be minimum at the values of the VEVs, they should satisfy the following tadpole equations.
\begin{eqnarray}
\mu_1 - \mu_{12} \frac{v_2}{v_1} + \lambda_1 v_1^2 + \frac{\lambda_{12}+\lambda_{12}'}{2} v_2^2 + \frac{\lambda_{1s}}{2} v_s^2 = 0 \, , \\
\mu_2 - \mu_{12} \frac{v_1}{v_2} + \lambda_2 v_2^2 + \frac{\lambda_{12}+\lambda_{12}'}{2} v_1^2 + \frac{\lambda_{2s}}{2} v_s^2 = 0 \, , \\
\mu_s + \frac{\lambda_{1s}}{2} v_1^2 + \frac{\lambda_{2s}}{2} v_2^2 + \lambda_s v_s^2 = 0 \, .
\end{eqnarray}
After the spontaneous breaking of the EW and $U(1)_X$ symmetries, we are left with three physical $CP$-even 
neutral Higgses, a charged Higgs, and a pseudoscalar Higgs. Following the restrictions given by the above minimization 
conditions, the mass matrix for the pseudo-scalars in $(\eta_1\ \eta_2 \ \eta_s)^T$ basis becomes
\begin{eqnarray}
M_A^2 = \frac{\mu_{12}}{v_1 v_2} \begin{pmatrix}
v_2^2 & -v_1 v_2 & \ 0 \\
-v_1 v_2 & v_1^2 & \ 0 \\
0 & 0 & \ 0
\end{pmatrix} \, .
\end{eqnarray}
It is evident from the mass matrix that two pseudoscalars remain massless after the diagonalization to their mass 
eigenstates. These two massless modes are eaten up by the two neutral gauge bosons, {\it viz.} $Z$ and $Z'$, to 
acquire masses. The remaining pseudoscalar is a physical state with a mass  $m_A=\sqrt{\dfrac{\mu_{12}}{v_1 v_2}v^2 }$, 
where $v=\sqrt{v_1^2+v_2^2} \simeq 246$~GeV.

It is worth noting that if the soft-breaking term was absent, i.e., $\mu_{12} = 0$ in the Lagrangian given in
Eq.~(\ref{eqn:lag}), all the pseudoscalars would have been massless. This is expected since, in the 
scalar sector of the Lagrangian, one can recover a global $U(1)$ symmetry, {\it viz.} 
$\phi \to e^{-i\theta Q}\phi$, where $\phi$ represents any of the scalars. This global symmetry remains intact even 
after both the SM and $U(1)_X$ gauge symmetries are spontaneously broken, leading to a massless 
physical scalar in the particle spectrum. The soft-breaking term is therefore needed to avoid this massless pseudoscalar.  

The mass matrix of the charged scalars in $(\phi_1^+\ \phi_2^+)^T$ basis is given by
\begin{eqnarray}
M_\pm^2 = \left(\frac{\mu_{12}}{v_1 v_2}-\frac{\lambda_{12}'}{2}\right) \begin{pmatrix}
v_2^2   & -v_1 v_2 \\
-v_1 v_2 & v_1^2   \\	\end{pmatrix} \, .
\end{eqnarray}
This $2\times2$ mass matrix can be easily diagonalized by rotating with an angle $\beta$, which is defined 
by the ratio of the VEVs of the two Higgs doublets given by $\tan\beta = \dfrac{v_2}{v_1}$. 
It should be noted that the same angle $\beta$ also diagonalizes the pseudoscalar mass matrix. One of 
the charged scalar is massless and corresponds to the charged Goldstone, which is eaten up by the $W^\pm$ gauge 
boson to get its mass. The remaining physical charged scalar is orthogonal to the massless one and is given by 
\begin{eqnarray}
H^\pm = -\sin\beta\,\phi_1^\pm +\cos\beta\,\phi_2^\pm
\end{eqnarray}
with mass $m_{H^\pm}=\sqrt{\left(\frac{\mu_{12}}{v_1 v_2}-\frac{\lambda_{12}'}{2}\right)v^2}$.

The $CP$-even scalar mass matrix  in the $(\rho_1 \,\, \rho_2 \,\, \rho_s)^T$ basis is given by
\begin{eqnarray}
M_H^2 = \begin{pmatrix}
2\lambda_1 v_1^2 + \mu_{12}\frac{v_2}{v_1}& (\lambda_{12}+\lambda'_{12})v_1 v_2 -\mu_{12} &~& \lambda_{1s}\,v_1 v_s \\
(\lambda_{12}+\lambda'_{12})v_1 v_2 -\mu_{12} & 2\lambda_2 v_2^2 + \mu_{12}\frac{v_1}{v_2} &~& \lambda_{2s}\,v_2 v_s \\
\lambda_{1s}\,v_1 v_s & \lambda_{2s}\,v_2 v_s &~& 2\lambda_s v_s^2
\end{pmatrix} \, , 
\end{eqnarray}
In general, the determinant of the mass matrix of $CP$-even scalar is nonzero, which tells us that there will be three
massive $CP$-even scalars after the symmetry breaking. We identify the three $CP$-even mass eigenstates as 
$h_1, h_2$, and $h_3$. They are linear combinations of the flavor states and can be written as
\begin{eqnarray}
h_i = Z_{ij}^h \,\, \rho_j \,\, ,
\end{eqnarray}
where $Z_{ij}^h$ represents the mixing matrix for the $CP$-even states.

For our analysis, we hereafter denote $h_1, h_2$, and $h_3$ as the physical eigenstates in ascending order of their masses. 
For simplicity, we restrict our choice on the parameters in the scalar sector such that the lowest 
mass eigenstate among all scalars will be the $125$~GeV scalar, identified as the SM Higgs boson observed at the experiments. 
As we do not consider a full analysis of the scalar sector in this work, it helps us to focus solely on the $Z'$ and heavy 
neutrinos of the model. The other two $CP$-even states are taken to be beyond $700$~GeV. 
As the properties of the lightest scalar must be similar to the SM Higgs boson, we choose the parameters such that 
$h_1$ belongs mainly to the first Higgs doublet $H_1$. In terms of the mixing matrix components $|Z_{11}^h|^2 \simeq 1$. 
This natural choice is easily achieved if the diagonal entries of mass matrix $M_H^2$ are much larger compared to the 
off-diagonal entries. This choice also suggests that $v_1\simeq v$, 
which implies that $\tan\beta \ll 1$. We discuss the choice of $\tan\beta$ further in Sec.~\ref{sec:lepqu}.
In this setup, the three diagonal entries are controlled by $v_1$, $\mu_{12}$, and $v_s$. So, the mass of the heavy scalars will  
be given~(to an approximation) by $m_{h_2} \simeq \sqrt{\mu_{12}\cot\beta}$, and 
$m_{h_3} \simeq \sqrt{2\lambda_s v_s^2}$. The mass for the charged scalar as well as the pseudoscalar will also be similar 
to the mass of $h_2$.
\subsection{Gauge kinetic mixing and masses of gauge bosons}\label{sec:GaugeMass}
The presence of two or more $U(1)$ gauge group in a theory allows us to write a gauge kinetic mixing term between the 
two $U(1)$ gauge bosons without spoiling the gauge invariance of the Lagrangian~\cite{Chankowski:2006jk}. The kinetic term for the 
gauge bosons in the Lagrangian, after including the gauge kinetic mixing, then becomes
\begin{eqnarray}
\lag \supset - \frac{1}{4} G^{a,\mu\nu} G_{\mu\nu}^a - \frac{1}{4} W^{b,\mu\nu} W_{\mu\nu}^b -\frac{1}{4} B^{\mu\nu} B_{\mu\nu} - \frac{1}{4} C^{\mu\nu} C_{\mu\nu} + \frac{1}{2} \tilde{g} B^{\mu\nu} C_{\mu\nu} \, , 
\end{eqnarray}
where $\tilde{g}$ is the kinetic mixing parameter. The following field redefinitions make the kinetic term diagonal with the desired coefficient
\begin{eqnarray}
B^\mu &=& B'^\mu + \dfrac{\tilde{g}}{\sqrt{1-\tilde{g}^2}} C'^\mu  \, ,  \\
C^\mu &=& \dfrac{1}{\sqrt{1-\tilde{g}^2}} C'^\mu   \, . \label{eqn:kinmix}
\end{eqnarray}
The field redefinition tells us than $\tilde{g}$ should be less than 1 for the fields to be real. This is usually referred to as the ``theoretical constraint' on $\tilde{g}$.  After achieving the correct form for the gauge kinetic term with the above field redefinitions, we can now 
try to write the mass terms of gauge bosons arising from the kinetic terms of the scalars,
\begin{eqnarray}
\mathcal{L}_{m,\rm gauge} &=& \left(D^\mu \langle H_1\rangle \right)^\dagger D_\mu \langle H_1\rangle + \left(D^\mu \langle H_2\rangle \right)^\dagger D_\mu \langle H_2\rangle + \left(D^\mu \langle S\rangle \right)^\dagger D_\mu \langle S \rangle   \, , 
\end{eqnarray}
where
\begin{eqnarray}
\langle H_1\rangle = \begin{pmatrix} \dfrac{v_1}{\sqrt 2} \\ 0\end{pmatrix},\qquad\qquad
\langle H_2\rangle = \begin{pmatrix} \dfrac{v_2}{\sqrt 2} \\ 0\end{pmatrix},\qquad\qquad
\langle S\rangle = \dfrac{v_s}{\sqrt 2}
\label{eqn:vevs2}
\end{eqnarray}
with the gauge covariant derivatives for the corresponding scalars defined as
\begin{eqnarray}
D_\mu^{(1)} &=& \partial_\mu -ig_2 \frac{\sigma^a}{2}W_\mu^a + i\frac{g_1}{2}B_\mu  \, , \\
D_\mu^{(2)} &=& \partial_\mu -ig_2\frac{\sigma^a}{2}W_\mu^a +i\frac{g_1}{2}B_\mu + i g_x q_x C_\mu  \, , \\
D_\mu^{(S)} &=& \partial_\mu -2ig_x q_x C_\mu  \, . 
\end{eqnarray}

The $U(1)_X$ charges of all the fields are proportional to $q_x$. In a gauge theory, the gauge coupling always comes
with the gauge charges, i.e., the constant that we will see is $g_x q_x$. This means that we can absorb $q_x$ in $g_x$. So, we will take $q_x=1$ henceforth. With the VEVs as defined in Eq.~(\ref{eqn:vevs2}), we get mass terms for the gauge bosons as follows:
\begin{eqnarray}
\mathcal{L}_{m,\rm gauge} &=&  \frac{1}{4}\begin{vmatrix} \begin{pmatrix}
g_2 W_\mu^3 - g_1 B_\mu & g_2(W_\mu^1 - iW_\mu^2)\\
g_2(W_\mu^1 + iW_\mu^2) & -g_2 W_\mu^3 - g_1 B_\mu
\end{pmatrix} \begin{pmatrix}
\dfrac{v_1}{\sqrt 2} \\ 0 
\end{pmatrix}
\end{vmatrix}^2 \nonumber\\
& & +\ \frac{1}{4}\begin{vmatrix} \begin{pmatrix}
g_2 W_\mu^3 - g_1 B_\mu - 2 g_x C_\mu & g_2(W_\mu^1 - iW_\mu^2)\nonumber\\
g_2(W_\mu^1 + iW_\mu^2) & -g_2 W_\mu^3 - g_1 B_\mu - 2g_x C_\mu
\end{pmatrix} \begin{pmatrix}
\dfrac{v_2}{\sqrt 2} \\ 0 
\end{pmatrix}
\end{vmatrix}^2  + 2 g_x^2 v_s^2 C_\mu C^\mu \nonumber\\
&=& \frac{1}{4}g_2^2 v^2 W^+_\mu W^{-\mu} + \frac{1}{8}v_1^2 \left|\left(g_2W^3_\mu - g_1 B'_\mu - \frac{g_1\tilde{g}}{\sqrt{1-\tilde{g}^2}} C'_\mu\right)\right|^2 \nonumber\\ 
& & +\ \frac{1}{8}v_2^2 \left|\left(g_2 W^3_\mu - g_1 B'_\mu - \frac{g_1\tilde{g}}{\sqrt{1-\tilde{g}^2}} C'_\mu - \frac{2g_x}{\sqrt{1-\tilde{g}^2}} C'_\mu \right)\right|^2 + \frac{2 g_x^2 v_s^2}{(1-\tilde{g}^2)}C'_\mu C'^\mu \nonumber  \, . 
\end{eqnarray}

From Eq.~(\ref{eqn:kinmix}), we see that $C'_\mu$ is always accompanied by the factor $\dfrac{1}{\sqrt{1-\tilde{g}^2}}$. Since the coupling $g_x$ always comes with $C_\mu$, and hence with $C'_\mu$, we may absorb this extra factor inside $g_x$. Also, from the above equation, we see that $\tilde{g}$ does not appear separately. Hence, without loss of generality, we do the following redefinitions in the coupling in order to get simplified expressions 
\begin{eqnarray}
g'_x = \frac{g_1\tilde{g}}{\sqrt{1-\tilde{g}^2}}  \, ,  \qquad g_x \to g_x\sqrt{1-\tilde{g}^2}  \, . \label{eqn:redef}
\end{eqnarray}
In Eq.~(\ref{eqn:redef}), the last redefinition means that we replace $g_x$ by $g_x\sqrt{1-\tilde g^2}$ in each place in the Lagrangian.
We should also note that there is no restriction on $g'_x$ from theoretical constraint even though we had restrictions on $\tilde g$. 

The mass matrix for the neutral gauge bosons, in the basis
of $\left(B'_\mu\  W^3_\mu\ C'_\mu\right)^T$ is given by 
\begin{eqnarray}
\!M^2 = \frac{1}{4}\!\begin{pmatrix}
g_1^2 v^2   & -g_1 g_2 v^2 &  g_1 \left(g'_x v^2 + 2 g_x v_2^2\right) \\
-g_1 g_2v^2 & g_2^2 v^2    & -g_2 \left(g'_x v^2 + 2 g_x v_2^2\right) \\
g_1 \left(g'_x v^2 + 2 g_x v_2^2\right) & ~~-g_2 \left(g'_x v^2 + 2g_x v_2^2\right) & ~{g'_x}^2 v^2 + 4 g_x g'_x v_2^2 + 4 g_x^2(v_2^2+ 4 v_s^2)
\end{pmatrix}.
\end{eqnarray}
The diagonalization of the mass matrix of the neutral gauge
bosons can be done in the following way,
\begin{enumerate}
	\item[(i)] First we rotate $W^3_\mu$ and $B'_\mu$ to get $A_\mu$ and $X_\mu$.
	\begin{eqnarray}
	\begin{pmatrix} A_\mu \\ X_\mu \\ C'_\mu \end{pmatrix} =
	\begin{pmatrix} \cos\theta_W & \sin\theta_W & 0\\ -\sin\theta_W & \cos\theta_W & 0 \\ 0 & 0 & 1\end{pmatrix}
	\begin{pmatrix} B'_\mu \\ W^3_\mu \\ C'_\mu\end{pmatrix}  \, , 
	\end{eqnarray}
	where $\tan\theta_W = \dfrac{g_1}{g_2}$.
	The mass term for neutral gauge boson then becomes
	\begin{eqnarray}
	\lag_{m,\text{gauge}} = \frac{1}{8} v_1^2 \left(g_z X_\mu - g'_x C'_\mu\right)^2 + \frac{1}{8} v_2^2 \big(g_z X_\mu - (g'_x+2g_x) C'_\mu\big)^2 + 2 g_x^2 v_s^2 C'_\mu C'^\mu  \, , 
	\end{eqnarray}
	where $g_z = \sqrt{{g_1}^2 + g_2^2}$. The above expression does not have any mass term for $A_\mu$. This means $A_\mu$ is massless, which can be identified as the photon. The angle $\theta_W$ can be identified as the Weinberg angle as we get in the SM. 
	\item[(ii)] Now, the mass matrix of $X_\mu$ and $C'_\mu$ is given by
	\begin{eqnarray}
	\widetilde{M}^2 = \frac{1}{4}\begin{pmatrix} g_z^2 v^2 & -g_z \left(g'_x v^2 + 2g_x v_2^2\right)\\ -g_z \left(g'_x v^2 + 2g_x v_2^2\right) & ~~{g'_x}^2 v^2 + 4 g_x g'_x v_2^2 + 4 g_x^2(v_2^2+ 4 v_s^2)
	\end{pmatrix} \, . 
	\end{eqnarray}
	The above mass matrix can be diagonalized by the orthogonal transformation between $X_\mu$ and $C'_\mu$ as follows
	\begin{eqnarray}
	\begin{pmatrix} Z_\mu \\ Z'_\mu \end{pmatrix} = \begin{pmatrix}  \cos\theta' & \sin\theta' \\ -\sin\theta' & \cos\theta' \end{pmatrix}
	\begin{pmatrix} X_\mu \\ C'_\mu \end{pmatrix}  \, , 
	\end{eqnarray}
	where	
\begin{equation}\label{tanthetap}	
\tan2\theta' = \dfrac{2g_z \left(g'_x v^2 + 2g_x v_2^2\right)}{{g'_x}^2 v^2 + 4 g_x g'_x v_2^2 + 4 g_x^2(v_2^2+ 4 v_s^2) 
- g_z^2 v^2} \, . 
\end{equation}	

After the diagonalization, the mass of the physical gauge bosons are
\begin{eqnarray}
M_{Z,Z'}^2 &=& \frac{1}{8}\Big[g_z^2 v^2 + {g'_x}^2 v^2 + 4 g_x g'_x v_2^2 + 4 g_x^2(v_2^2+ 4 v_s^2)\Big] \\
& &  \mp  \frac{1}{8}\sqrt{\Big({g'_x}^2 v^2 + 4 g_x g'_x v_2^2 + 4 g_x^2(v_2^2+ 4 v_s^2) - g_z^2 v^2\Big)^2 + 4 g_z^2 \Big(g'_x v^2 + 2 g_x v_2^2\Big)^2}  \nonumber \, , 
\end{eqnarray}
and the final mixing matrix becomes
\begin{eqnarray}
\!\!\!\!\begin{pmatrix} B_\mu \\ W_\mu^3 \\ C_\mu \end{pmatrix}
\!&=&\! \begin{pmatrix} \cos\theta_W & ~-\sin\theta_W\cos\theta'  + \dfrac{g_x'}{g_1}\sin\theta' & ~\sin\theta_W \sin\theta' + \dfrac{g_x'}{g_1}\cos\theta'\\ \sin\theta_W & \cos\theta_W \cos\theta' & -\cos\theta_W \sin\theta' \\ 0 &  \sin\theta'\dfrac{\sqrt{g_1^2 + {g_x'}^2}}{g_1}  & \cos\theta'\dfrac{\sqrt{g_1^2 + {g_x'}^2}}{g_1} \end{pmatrix}\!\! \begin{pmatrix} A_\mu \\ Z_\mu \\ Z'_\mu\end{pmatrix}. 
\end{eqnarray}
\end{enumerate}

Note that the mixing between the $Z$ and $Z'$ needs to be quite small such that it does not modify the $Z$ boson couplings with 
the SM fields. In order to study the parameters that would be most relevant in establishing the $Z$-$Z'$ mixing, we look at 
Eq.~(\ref{tanthetap}) in more detail. We find that the kinetic mixing dictates that the coefficient $g_x'$ appears with the SM VEV
while the $U(1)_X$ coupling $g_x$ appears with the VEV of the second scalar doublet in the numerator of Eq.~(\ref{tanthetap}).  
Assuming that the kinetic mixing coefficient and the $U(1)_X$ gauge coupling are of the same order, one can approximate 
Eq.~(\ref{tanthetap}) depending on the choice of $\tan\beta$. Note that for $\tan\beta \ll 1$, i.e., $v_1 \simeq v$,
 the dominant term in the numerator becomes proportional to $g_z \, g_x' \, v_1^2$, while 
 for $\tan\beta \gg 1$, i.e.,  $v_2 \simeq v$, the dependence is on $g_z (2 g_x + g_x' ) v_2^2$. 
The denominator can be easily approximated to a form $(M_{Z'}^2-M_Z^2)$ in either case, provided $v_s \gg v_1, v_2$. 
Thus depending on the choice of $\tan\beta$, we expect the mixing angle to vary for different ranges of $g_x'$ and $g_x$ values.

\begin{figure}[h!] \vspace{-4mm}
\includegraphics[width=0.7\textwidth]{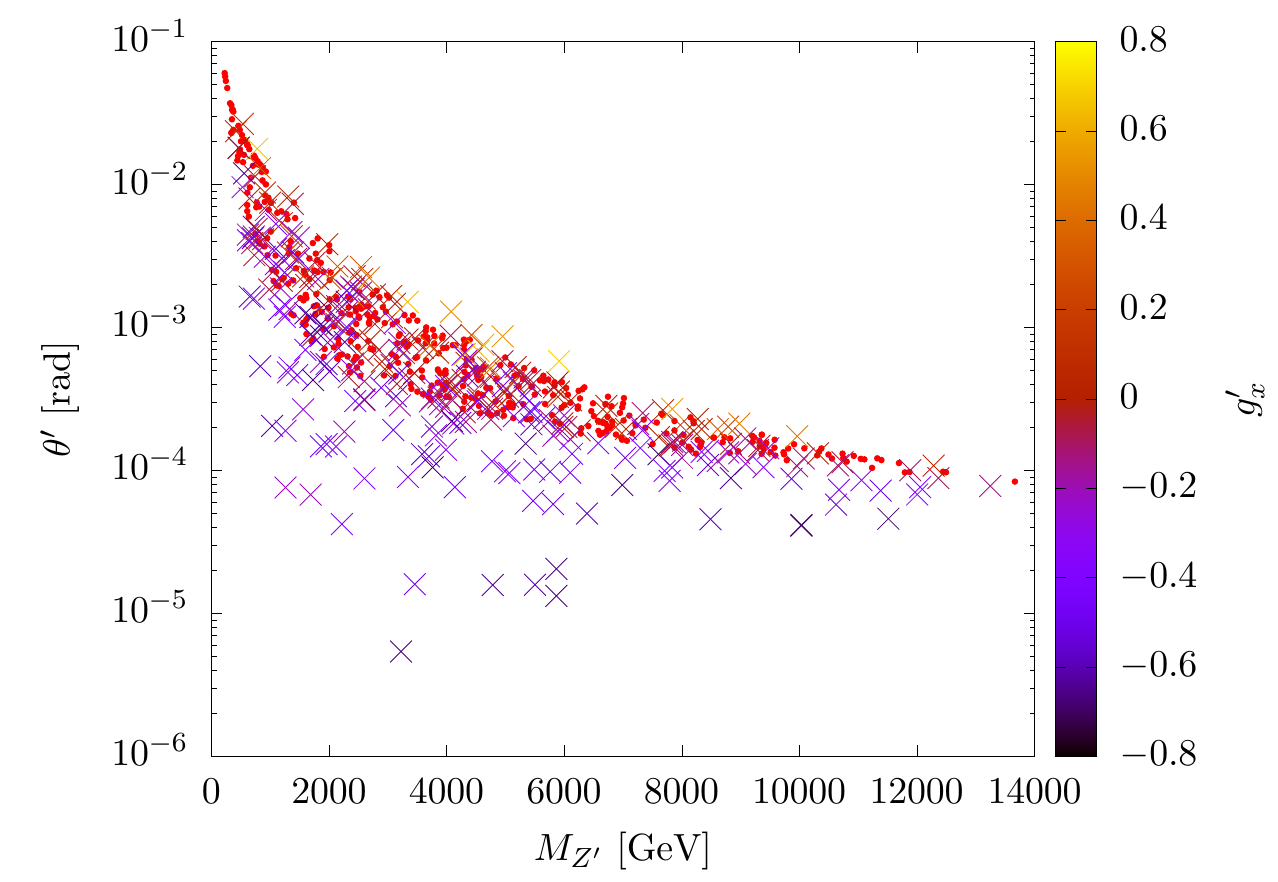}
\caption{$(M_{Z'},\theta')$ along with the gauge kinetic mixing $g'_x$ on the color bar. Red points refer to $g'_x=0$ while cross symbols ``$\times$'' indicate the nonvanishing $g'_x$. Here the scan is carried out for parameter values in the range
$1\leq v_s\leq 10$~TeV, $0.1\leq g_x\leq 0.7$, and $-1\leq g'_x\leq 1$.}
\label{Fig:MZp-thetap-gt}
\end{figure}
To highlight the case where $\tan\beta > 1$, i.e., $v_2 > v_1$, we scan over a range of values for $g_x'$ and $g_x$ 
as well as $v_s$ for $ 1 < \tan\beta < 60$ and calculate the mixing angle $\theta'$. In Fig.~\ref{Fig:MZp-thetap-gt}, we show the dependence of the 
$Z$-$Z'$ mixing angle $(\theta')$, as a function of  $M_{Z'}$ along with its dependence on the variation of the gauge 
kinetic mixing $(g'_x)$. Note that for large values of $M_{Z'} > 1$ TeV the denominator term is significantly large and therefore 
the mixing angle is naturally small. However the numerator in Eq.~(\ref{tanthetap}) is proportional to $g_z (2 g_x + g_x' ) v_2^2$ 
for $\tan\beta \gg 1$ and we find that even with the kinetic mixing vanishing, the mixing angle has values larger than 
$\mathcal{O}(10^{-2})$ for $M_{Z'} < 750$~GeV. This is expected as the denominator $(M_{Z'}^2-M_Z^2)$ becomes smaller, 
while $g_x$ is nonvanishing and constrained by the $Z'$ mass. This gives an interesting result that, even with vanishing 
kinetic mixing, if the $Z'$ gets a part of its mass from the scalar doublet, it leads to a substantial $Z$-$Z'$ mixing, which would 
disfavor the parameter space due to strong constraints from $Z$ boson measurements. However it is still possible to obtain 
small $\theta' < 10^{-3}$ for the light $Z'$ case, provided there is a cancellation in the numerator term $\propto (2 g_x + g_x' )$. 
These are the points highlighted in the figure with crosses~($\times$) corresponding to {\it negative} values of $g_x'$. 
Thus it is possible to 
obtain small $Z$-$Z'$ mixing compatible with $Z$ boson data even for $\tan\beta \gg 1$. The stronger constraint on such a 
scenario however comes from Higgs data and perturbativity arguments, which we discuss later along with the more favorable 
choice of parameter space where $\tan\beta \ll 1$.  

\subsection{Masses and mixing of the charged lepton and quarks}\label{sec:lepqu}
The Lagrangian responsible for the masses and the mixing of
leptons and quarks is essentially the Yukawa terms.
\begin{eqnarray}
\lag \supset -Y_l^{ij} \bar l_{Li} H_1^\mathcal{C} e_{Rj} - Y_d^{ij} \bar q_{Li} H_1^\mathcal{C} d_{Rj} - Y_u^{ij} \bar l_{Li} H_1 u_{Rj}  + {\rm H.c.}
\end{eqnarray}
The masses and the mixing can be arranged in the same way as it is done in the SM. The only difference is that the 
mass of the SM fermions are proportional to the VEV of $H_1$, $v_1$. So, in order to achieve the correct mass, 
we need to choose Yukawa couplings $Y_f = \frac{Y_f^{\rm SM}}{\cos\beta}$, where $Y_f^{\rm SM}=\dfrac{\sqrt{2}m_f}{v}$ 
is the value of the respective Yukawa couplings in the SM. This choice also ensures that the 
Cabibo-Kobayashi-Maskawa~(CKM) matrix remains the same as the SM. With this choice, we tabulate the 
couplings of the fermions to the scalars, namely $h_i \, (i=1,2,3)$, $A$ and $H^\pm$, in Table~\ref{tab:hffCoupling}.
In order to maintain perturbativity of all the couplings, we need to keep these coupling below $\sqrt{4\pi}$.  
From the table, it is clear that the natural choice for $\tan\beta$ is smaller values. The strongest constraint 
from perturbativity comes from the top quark since it is the heaviest fermion in the SM. In the case of top 
quark, $\sqrt{2}\frac{m_f}{v} \simeq 1$. Hence, if we take $Z^h_{i1}\simeq 1$, $\tan\beta$ should be such that 
$\cos\beta > \frac{1}{\sqrt{4\pi}}$ from perturbativity consideration. This gives, although an approximate one, 
an upper bound of $\tan\beta < 3$. With this bound in mind we shall restrict our study to values of $\tan\beta < 1$
for further analysis. Recall that for any value of $\tan\beta > 1$, there is significant increase in the couplings of the 
SM fermions with the scalars in the model. A critical scrutiny of its implications and phenomenology for the 
scalar sector in the current model is left for future work and we focus on the $Z'$ signal in this work.  
\begin{table}[hb!]
\begin{tabular}{|c|c|c|c|}
\hline
Couplings for   & ${h_i-f-\bar f}$ & $A-f-\bar f$ & $H^\pm-f-\bar{f'}$ \\
\hline
 $g_f$ & $Y_f^{\rm SM}\dfrac{Z_{i1}^h}{\cos\beta}$ & $Y_f^{\rm SM}\tan\beta$ & $Y_f^{\rm SM} \tan\beta$\\
\hline
\end{tabular}
\caption{The coupling of the fermions with different scalars of the model.}
\label{tab:hffCoupling}
\end{table}
\vspace{-8mm}
\subsection{Masses of neutrinos}
In this model, we give Majorana masses to the neutrinos via inverse seesaw 
mechanism~\cite{Mohapatra:1986bd, Nandi:1985uh, Mohapatra:1986aw}. We rewrite the relevant 
part of the Lagrangian below.
\begin{eqnarray}
\lag &\supset& -\ Y_\nu\,\overline{l_L} H_2 N_R
 - Y_R S \overline N_R N_R^C - Y_L S \overline N_L N_L^C - \hat{M}_N\overline{N}_L N_R + \text{h.c.}
\label{eqn:lag-neutrino}
\end{eqnarray}
We have added three generations of sterile neutrinos~($N_R^i$ and $N_L^i$) corresponding to the three generations of fermion in 
the SM, which renders all the Yukawa couplings~($Y_\nu$, $Y_L$ and $Y_R$) as $3\times3$ matrices. Note that 
the two chiral states $N_R$ and $N_L$ combine to form a vectorlike fermion~($\hat{N}$), which is a singlet under 
SM and carries the same $U(1)_X$ charge as its chiral components. After symmetry breaking, 
the mass term for the neutrinos are given by
\begin{eqnarray}
\lag_\nu^\text{mass} = -\frac{v_2}{\sqrt 2}Y_\nu \overline{\nu}_L N_R - \frac{v_s}{\sqrt 2}Y_R\overline{N_R^C} N_R - \hat{M}_N \overline{N}_L N_R - \frac{v_s}{\sqrt 2}Y_L\overline{N_L^C} N_L + {\rm H.c.}
\end{eqnarray}
The mass matrix in $\left(\nu_L\ N_R^C\ N_L \right)^T$ basis is given by
\begin{equation} 
{\cal M}_{\nu} = \left( 
\begin{array}{ccc}
0 &m_D^T  &0\\ 
m_D  &m_R  &\hat{M}_N\\ 
0 &{\hat{M}_{N}}^{T} &m_L \end{array}
\right) , 
 \end{equation}
 where $m_D= v_2 Y_\nu/\sqrt{2}$, $m_R=\sqrt{2} v_s Y_R$, and $m_L=\sqrt{2} v_s Y_L$. Also, $m_L$ 
 and $m_R$ are naturally small due to the so-called 't~Hooft criteria~\cite{tHooft:1979rat}. Indeed, in the limit 
 $m_{L,R}\ \rightarrow 0$, the lepton number is restored as a conserved symmetry.

As mentioned above, $m_L,m_R \ll m_D,\hat{M}_N$, thus the neutrino masses can be given, with
a very good approximation, by
\begin{eqnarray}
m_{\nu_\ell} & \simeq & \frac{m^2_D\, m_L}{\hat{M}_N^2+m_D^2},\label{mnul}\\
m_{\nu_{H,H'}} & \simeq & \frac{1}{2}\left(\frac{\hat{M}_N^2\,m_L}{\hat{M}_N^2+m_D^2}+ m_R\right) \mp \sqrt{\hat{M}_N^2+m_D^2}\,.
\end{eqnarray}
It is worth mentioning that, in this scenario, the neutrino Yukawa coupling $Y_\nu$,
can be of order ${\cal O}(0.1)$ and the large scale $\hat{M}_N$ can lie in the range of a few hundred~GeV--TeV. 
This is because the suppression factor needed to account for light neutrino masses are played by the naturally 
small parameters $m_L$ instead of the Yukawa coupling $Y_\nu$. Such a large Yukawa coupling plays a crucial 
role for producing these heavy neutrinos~(which are complete SM isosinglets) at experiments directly through SM 
mediators and helps in testing these type of models and probing the heavy neutrino physics at colliders~(some examples
as in Refs.~\cite{Das:2012ze,Dev:2013wba,Das:2014jxa,Deppisch:2015qwa,Mondal:2016kof}). 
Indeed, if $Y_\nu \sim{\cal O}(0.1)$, $\hat{M}_N \sim 1$~TeV, and $m_L \sim  {\cal
O}(10^{-4})$~GeV, then an order of ${\cal O}(0.01)$~eV neutrino mass can be obtained.

The light neutrino mass matrix in Eq.~(\ref{mnul}) must be diagonalized by the physical neutrino
mixing matrix $U_{\rm PMNS}$~\cite{Zyla:2020zbs}, {\it i.e.}, %
\begin{equation} %
U_{\rm PMNS}^T m_{\nu_\ell} U_{\rm PMNS} = m_{\nu_\ell}^{\rm diag} \equiv
{\text{ diag}}\{m_{\nu_e}, m_{\nu_\mu}, m_{\nu_\tau}\}.%
 \end{equation}
Thus, one can easily show that the Dirac neutrino mass
matrix can be defined as :%
\begin{equation}  %
m_D=U_{\rm PMNS}\, \sqrt{m_{\nu_\ell}^{\rm diag}}\, R\, \sqrt{m^{-1}_L}\, \hat{M}_N, %
\label{mD2}
 \end{equation} %
where $R$ is an arbitrary orthogonal matrix. Accordingly, the $(9\times 9)$ neutrino mass matrix ${\cal M}_\nu$ can be diagonalized by ${\cal N}$,
{\it i.e.}, ${\cal N}^T {\cal M}_\nu \,{\cal N} = {\cal
M}_\nu^{ \rm diag}$, which is given by~\cite{Dev:2009aw}%
\begin{equation} \label{VCKM2}
 {\cal N}=\left(%
\begin{array}{cc}
  {\cal N}_{3\times 3} & {\cal N}_{3\times6}\\
  {\cal N}_{6\times 3} & {\cal N}_{6\times6}  \\
\end{array}%
\right),%
 \end{equation}%
where  %
\begin{equation} %
{\cal N}_{3\times3} \simeq \left(1-\frac{1}{2} F^2 \right)
U_{\rm PMNS}, ~~~~~  {\cal N}_{3\times6}=\left({\bf 0}_{3\times3},F \right) {\cal N}_{6\times6},~~~~~~  F = m_D \hat{M}^{-1}_N.%
 \label{eq:vln}
 \end{equation}%
It is clear that the deviation of a nonunitary matrix ${\cal N}_{3\times 3}$ from the standard $U_{\rm PMNS}$ is
measured by the size of $\frac{1}{2} F^2$. Also, the muon $g-2$ anomaly and the lepton flavor violating processes can be affected 
by the $F$ size~\cite{Abdallah:2011ew}. Consequently, that imposes upper bounds on $F$ entries to be 
small~\cite{Antusch:2006vwa,Malinsky:2009df,Ibarra:2011xn}, which is automatically satisfied in our model due 
to the smallness of $v_2$~(i.e., $v_1\simeq v$). 


In normal hierarchy scenario, i.e., assuming $m_{\nu_1} < m_{\nu_2} < m_{\nu_3}$, the two mass square differences determined from the 
oscillation data~\cite{deSalas:2017kay} is given by $\Delta m_{21}^2 = (7.05 - 8.24) \times 10^{-5}$~eV$^2$ and 
$\Delta m_{31}^2 = (2.334 - 2.524) \times 10^{-3}$~eV$^2$. Therefore, there are  at least two nonzero~$m_{\nu_i}$. 
Assuming the lightest neutrino to be massless, we get $m_{\nu_i} \simeq (0, 8.66 \times 10^{-3}, 0.05)$~eV. 

\begin{figure}[t!]
\begin{center}
\includegraphics[width=0.45\textwidth]{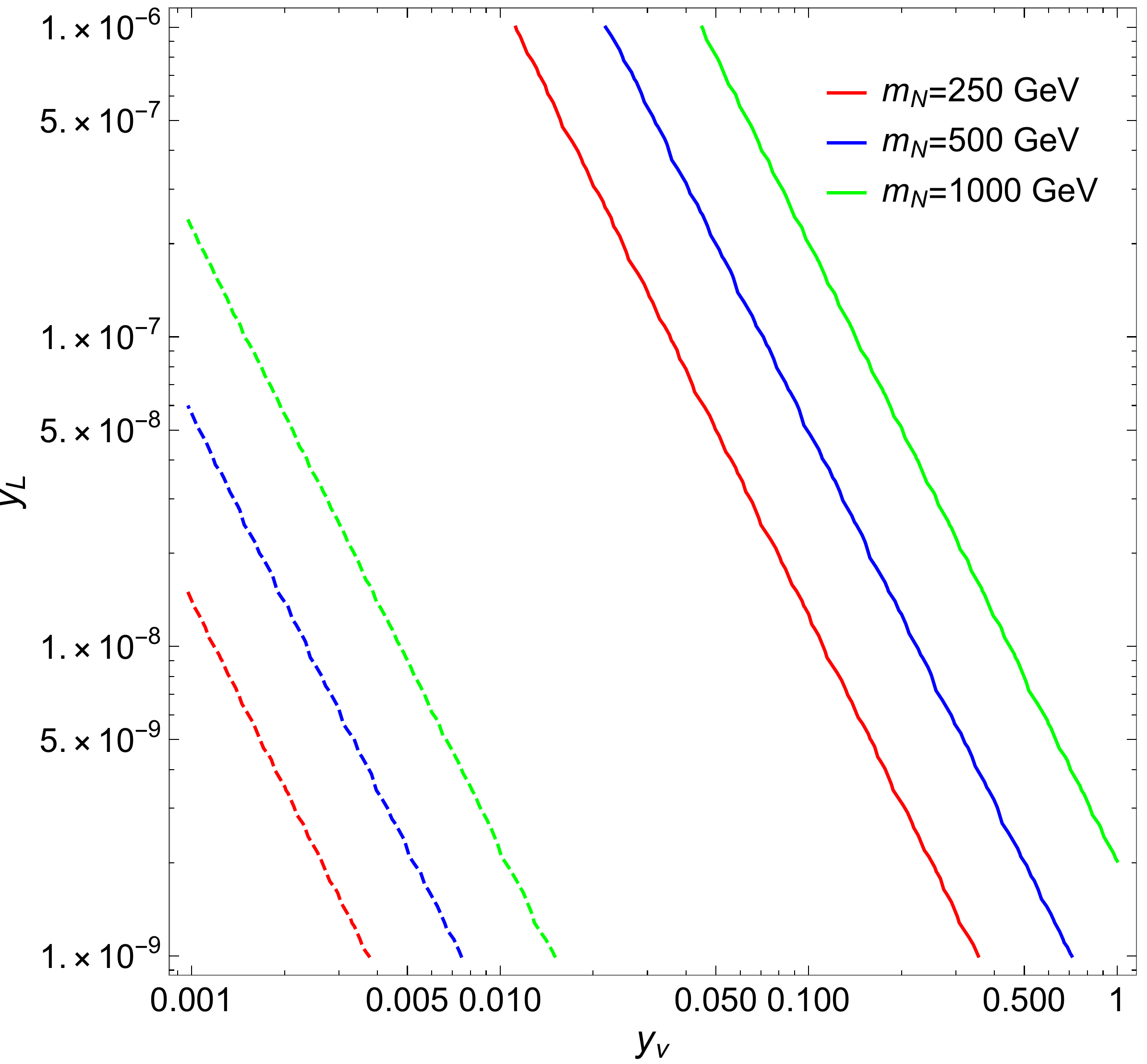} \hspace*{0.25in}
\includegraphics[width=0.45\textwidth]{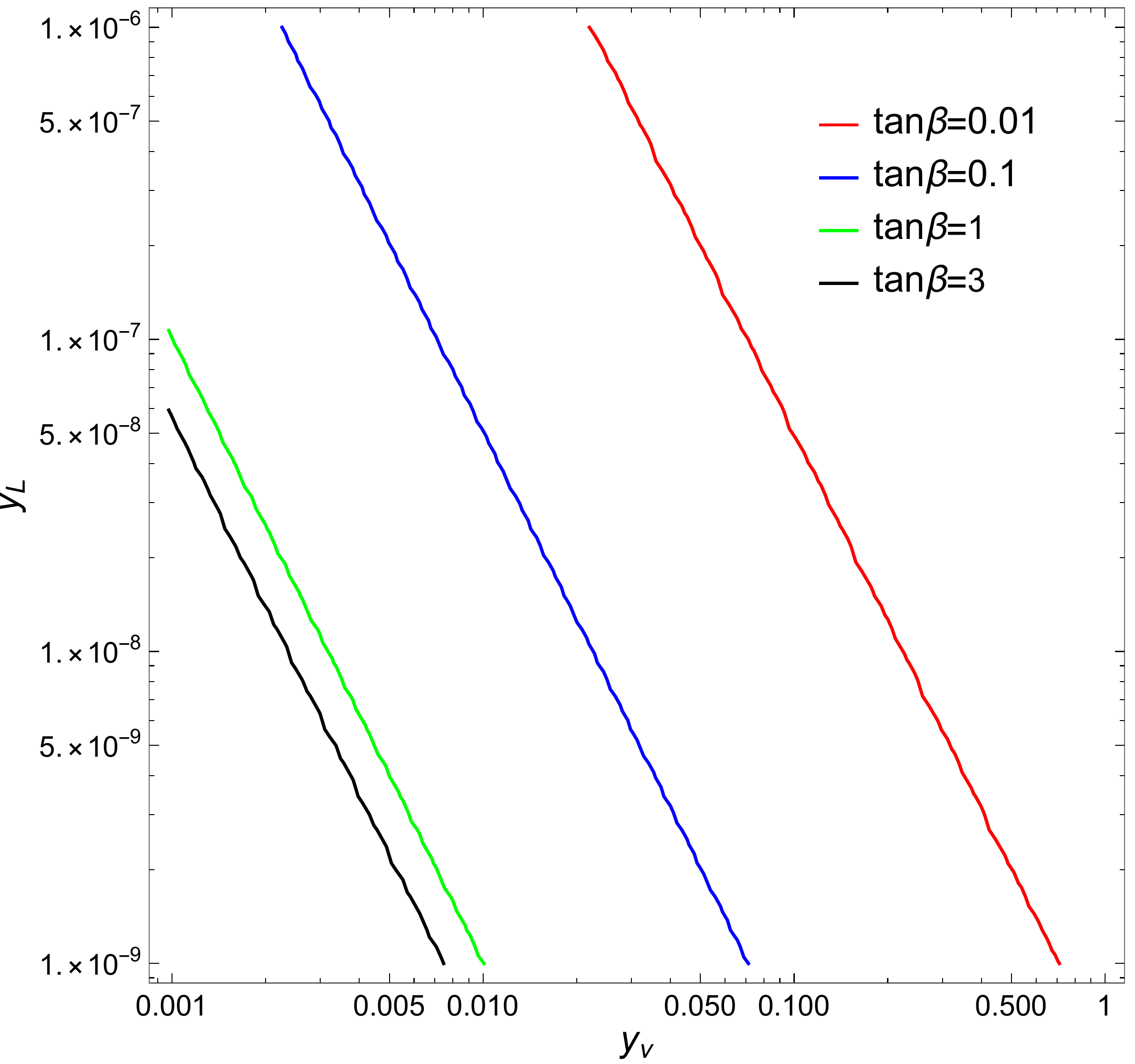}
\end{center}
\caption{$(y_\nu,y_L)$ plane in which the all curves satisfy the central values of the difference of  neutrino masses squared~($\Delta m_{21}^2,~\Delta m_{31}^2 $).}
\label{Fig:Yv-YL-mv2}
\end{figure}

For simplicity, we assumed $Y_\nu, \hat{M}_N$ to be diagonal and $Y_\nu^{ii}=y_\nu, \hat{M}_N^{ii}=m_N,~i=1,2,3$, and $Y_R=0$. 
Also, we defined $y_L=Y_L^{22}=Y_L^{33}\sqrt{\Delta m_{21}^2 /\Delta m_{31}^2 }$. In Fig.~\ref{Fig:Yv-YL-mv2}, 
we show the allowed $y_\nu$ and $y_L$ ranges to satisfy the central values of the difference of  
neutrino masses squared~($\Delta m_{21}^2,~\Delta m_{31}^2 $) for three different values of $m_N=250, 
500, 1000~{\rm GeV}$~(left panel), where the solid~(dashed) curves refer to $\tan\beta=0.01(3)$ and in the 
right panel we show the same but for different values of $\tan\beta=0.01,0.1,1,3$ with $m_N$ fixed at 500~GeV.
\section{Experimental Constraints} \label{sec:constraints}
The extension to the SM considered in this model affects the three sectors of the SM, {\it viz.} (i) scalar sector, (ii) neutrino sector, and (iii) neutral gauge boson sector.  We therefore need to focus on each of these to evaluate the experimental constraints  
that affect the parameter space of the model. 

\subsection{Properties of the $Z$ boson}
Due to the mixing of the gauge bosons, the coupling of $Z$ boson to SM particles gets modified with 
respect to that of the SM. As a result, the total decay width of the $Z$ boson as well as its partial decay width 
to light neutrinos~(which mix with the heavy neutrinos) is also modified. The modification in all the couplings besides the 
neutrinos is approximately proportional to $\sin\theta'$~(the $Z$-$Z'$ mixing parameter). The $Z$ boson properties have been 
measured at the Large Electron-Positron collider (LEP) with great precision and any changes to its  decay properties result in the limit for 
$\theta' \lesssim 10^{-3}$~\cite{Zyla:2020zbs}. This restriction puts a very strong constraint on the parameter 
space~[viz. Eq.~(\ref{tanthetap})].  In order to respect the constraints arising from the properties of the $Z$ boson, we 
choose the parameters of our model such that $\theta' < 10^{-3}$ is satisfied. As one can see from 
Eq.~(\ref{tanthetap}), the value of $\theta'$ depends on the coupling $g_x$ and gauge kinetic mixing $g_x'$ 
as well as the value of the EW VEVs, viz. $\tan\beta = \dfrac{v_2}{v_1}$ and $v_s$. 
As pointed out earlier, for high values of $v_s$ leading to $M_{Z'} > 1$ TeV, this bound is easily satisfied. Again, for 
$\tan\beta > 1$, we already discussed the regions of parameter space that is allowed for lower mass of $Z'$ 
in the concluding part of Sec. \ref{sec:GaugeMass}. Our interest lies in the parameter space with the more compatible 
choice of $\tan\beta < 1$ which allows a lighter $Z'$. 

\begin{figure}
\centering
\includegraphics[width=0.55\textwidth]{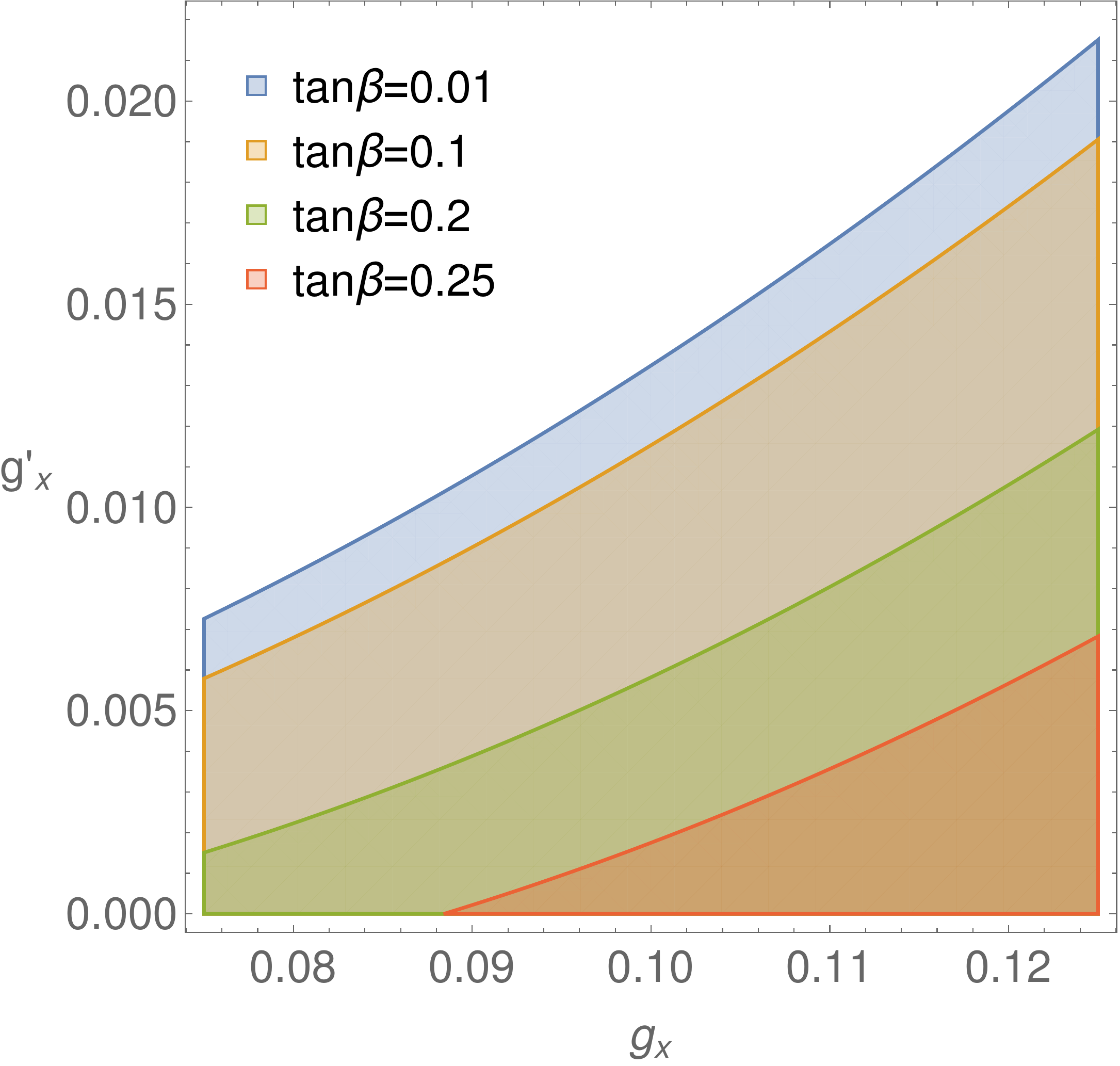} 
\caption{Illustration of allowed region satisfying
$\theta^{'} \le 10^{-3}$ in $g_x$-$g_x'$ plane for
$v_s=2$~TeV for four different value of $\tan\beta$.} \label{fig:thp}
\end{figure}
In Fig.~\ref{fig:thp}, we show the allowed region in the
$g_x$-$g_x'$ plane for $v_s=2$~TeV for different values of $\tan\beta$ less than one. The range of 
$g_x$ is chosen such that the mass of $Z'$ remains within $200$--$500$~GeV.  As the mass of the $Z'$ is approximated by 
$M_{Z'} \sim\!2 \, g_x v_s$, the value of $M_{Z'}$ within a certain range allows us to fix $g_x$ appropriately for a fixed 
value of $v_s$. As pointed out earlier, for $\tan\beta \ll 1$ we have the numerator in Eq.~(\ref{tanthetap}) proportional to 
the product of $g_x'$ and $v_1^2$. Thus, for $g_x'=0$, even a $g_x \sim \mathcal{O}(1)$ is allowed for the $U(1)_X$ gauge coupling. Thus, substantially large values of $g_x$ is allowed even when the $Z'$ mass lies 
between $200$--$500$~GeV, restricted 
only by the choice of $v_s$. This possibility leads us to the choice of the coupling which allows the $Z'$ to decay dominantly 
to a pair of the heavy neutrinos~(when kinematically allowed) while all other modes are suppressed. We will see that this 
also helps us evade existing collider limits on light $Z'$.
\subsection{Constraints from {\tt HiggsSignals} and {\tt HiggsBounds}}
The introduction of another Higgs doublet and singlet
modifies the scalar sector. The modifications are of the
following two forms. 
\begin{itemize}
\item Due to the mixing between scalars, the production and
branching fraction of the observed $125$~GeV scalar gets
modified with respect to the SM Higgs. These properties are
measured in terms of signal strength of Higgs which gives
constraint on the parameters~\cite{Aaboud:2018xdt,Sirunyan:2018ouh,Sirunyan:2017khh,Aaboud:2018gay,ATLAS:2020wny,Sirunyan:2018egh,CMS:2021ugl,Aad:2019lpq}.
\item The model predicts heavy scalars which may be observed
at the LHC. However, the LHC did not observed any new scalar
other the $125$~GeV one. This gives another constraints on the
production of any new scalars.   
\end{itemize}

Note that the choice of small $\tan\beta$ leads to suppressed couplings 
of charged scalars and pseudoscalar to the fermions as can be seen from the 
couplings shown in Table \ref{tab:hffCoupling}. As a result, the production of these scalars 
at a collider are significantly suppressed. This helps us to evade any bounds
coming from the nonobservation of such scalars at the LHC. 
However, the coupling of $CP$-even scalars~($h_i$) to the fermions are not all
suppressed due to the small values of $\tan\beta$. These couplings are mainly dictated by 
the entries in the $CP$-even scalar mixing matrix given by $Z^h_{i1}$. Since we
demand that the $125$~GeV scalar belongs mainly to the $H_1$ doublet,
we restrict ourselves to $Z^h_{11}\simeq1$ and $Z^h_{21},Z^h_{31} \ll 1$. This leads to the
suppressed production rates for the two heavy $CP$-even scalars while 
ensuring that the properties of the $125$~GeV scalar~($h_1$) resembles the SM Higgs. 

Although we do not explore the Higgs sector of the model in this article, we need to ensure that the 
parameter choice for the scalar sector satisfies all relevant constraints including that of the 
observed Higgs boson mass and its decay probabilities. To achieve this we use the publicly available 
packages {\tt HiggsSignals}~\cite{Bechtle:2013xfa} and {\tt HiggsBounds}~\cite{Bechtle:2008jh,Bechtle:2011sb}
in our scan of the parameter space to check for compatible points. 
These two packages incorporate the constraints of Higgs signal strength of the $125$~GeV scalar and 
also check the existing limits on the heavy scalars~(at 95\% C.L.).  
We shall henceforth fix the scalar sector parameters and masses consistent with relevant experimental constraints.
The parameter choices and the corresponding scalar masses are shown in Table~\ref{tab:scalar}.
\begin{table}[h]
\centering
\begin{tabular}{|c|c|c|c|c|c|c|c||c|c|c|c|}\hline
 ${\lambda_{1}}$ & $\lambda_{2}$ & $\lambda_{3}$ & $\lambda_{4}$ & $\lambda_{1s}$ & $\lambda_{2s}$  & $\mu_{12}$ (GeV$^2$) & $\tan\beta$ & $m_{h_{1}}$(GeV) & $m_{h_{2}}$ (TeV) & $m_{H^\pm}$ (GeV) & $m_{A}$ (TeV) \\
\hline
 0.1289 & 1.0 & 0.005 & 0.005 & 0.0 & $-$0.5  & $10^{4}$ & 0.01& 125.0 & 1.0  & 999.9  & 1.0  \\ \hline 
\end{tabular}
\caption{Scalar sector parameters and masses consistent with all experimental constraints.}
\label{tab:scalar}
\end{table}

The only parameter that we do vary in the scalar sector when we fix the benchmark points for our analysis 
would be the singlet VEV $v_s$ and the corresponding quartic term coefficient $\lambda_s$, which 
will affect the $Z'$ and $h_3$ masses. 

\subsection{Search for new $Z'$ gauge boson}
The phenomenology of $Z'$ in the model is quite different from that of the more traditional 
$U(1)$ extensions. In the absence of gauge kinetic mixing, the coupling of $Z'$ to the SM fermions 
gets modified by an additive factor proportional to $\sin\theta'$, which has to be small to be consistent 
with the measurement of $Z$ boson properties. However, the introduction of kinetic mixing parametrized by 
$g_x'$, we have an additional part in coupling, which is proportional to $g_x' \cos\theta'$. We have listed the
expression for the coupling of the $Z'$ with the matter fields of the model in the Appendix for reference.

As none of the SM fields are charged under the new $U(1)$, the $Z'$ couples to the SM charged fermions only via 
the $Z$-$Z'$ mixing. For $\tan\beta > 1$ we found that the mixing angle was dependent on both $g_x$ and $g_x'$. 
A small $\theta' \lesssim 10^{-3}$ for $M_{Z'}$ in the range of $200$--$500$~GeV required a cancellation such 
that $g_x' \simeq - 2 g_x$. However, this choice would imply that the coupling of the $Z'$ with the SM fermions and the 
new heavy neutrinos would have somewhat similar strength. Thus, in order to have substantial production cross section, 
one also gets a substantial branching fraction of the $Z'$ decay into SM fermions. For a light $Z'$, the strongest 
constraint from the LHC comes from its decay into the dilepton channel~\cite{Aaboud:2017buh}. Evaluating this limit for 
the case $\tan\beta > 1$, puts a strong limit on the values of $g_x'  \,\, {\rm and} \,\, g_x \sim 10^{-3}$.  Thus the promising search 
channel, when $\tan\beta > 1$, still remains the dilepton mode, even with the heavy neutrino decay modes available 
for the $Z'$. In contrast, when we consider the more favorable option of $\tan\beta < 1$, we find that the constraint 
on $\theta'$ is much more easily satisfied by suppressing the kinetic mixing parameter $g_x'$~(even for light $Z'$) 
while the decay modes of the gauge boson can be significantly tilted in favor of the new neutral fermions in the 
particle spectrum. However, a too suppressed $g_x'$ would also suppress the production cross section of the $Z'$ 
at the LHC, as can be seen by looking at its coupling with the SM quarks~(see the Appendix). We would therefore like to find a 
region of parameter space where the gauge boson is produced at the LHC and leaves an observable imprint in final 
states still allowed by the LHC data.

\begin{figure}[ht!]
\centering
\includegraphics[width=0.4\textwidth]{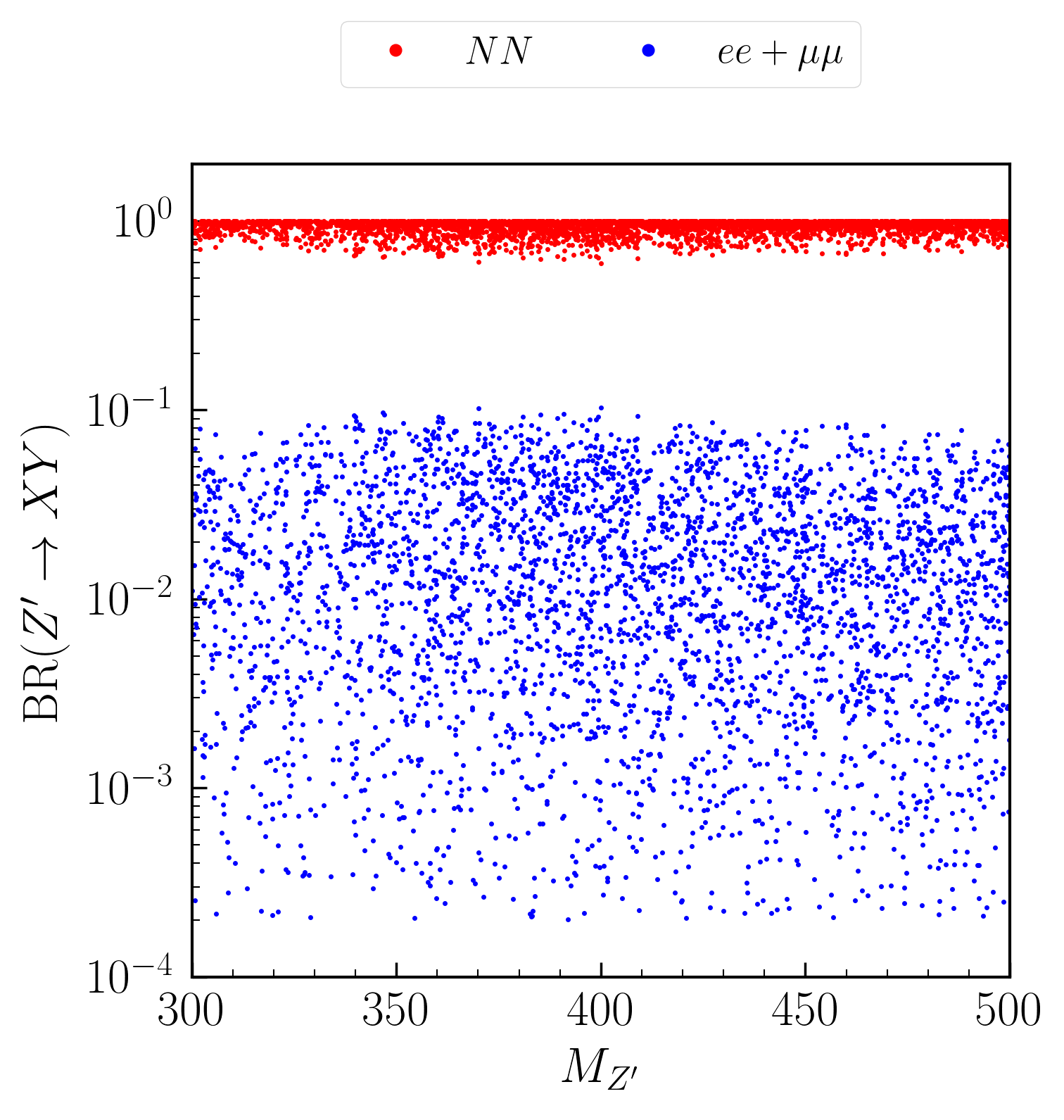} \hspace*{0.25in}
\includegraphics[width=0.4\textwidth]{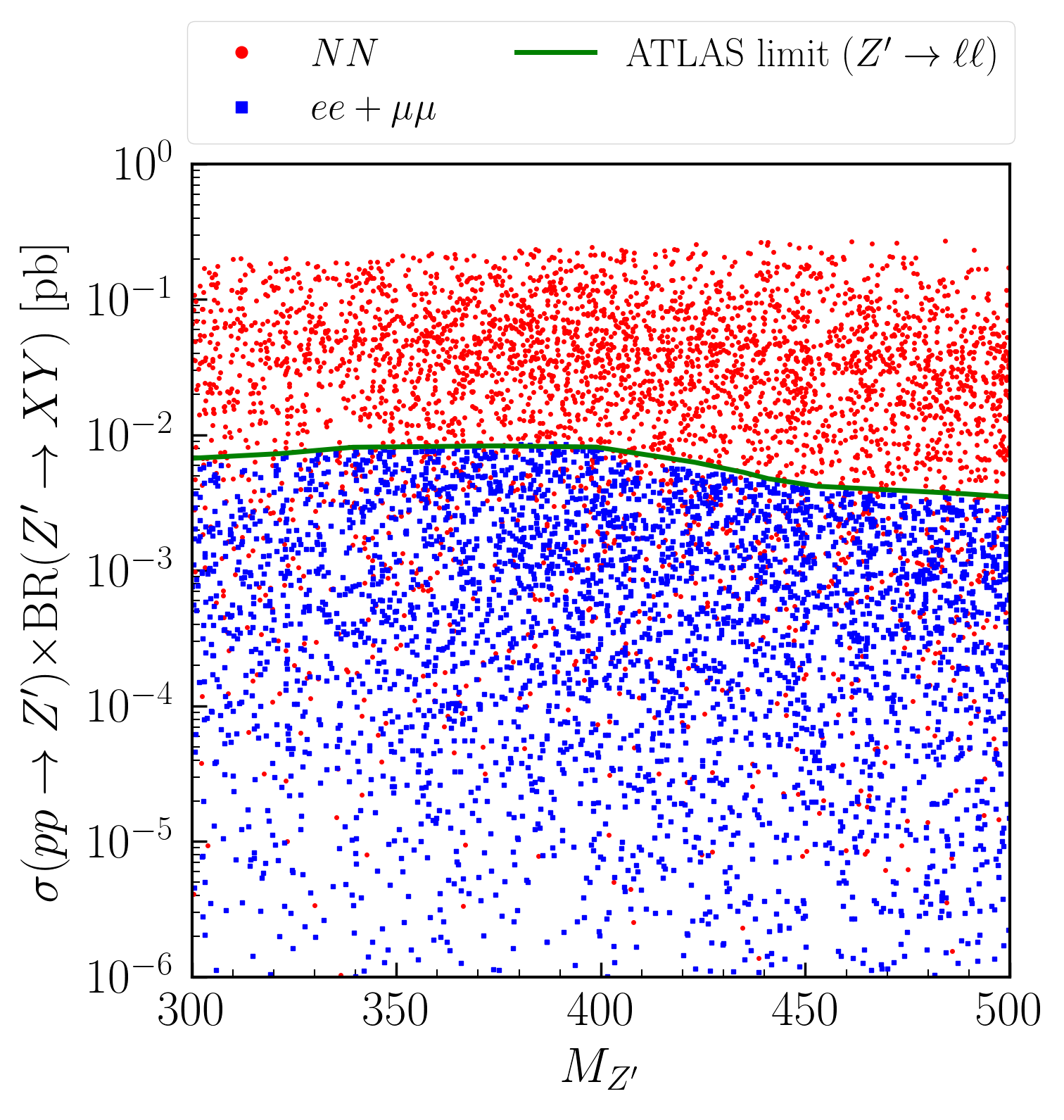} 
\caption{Left: scatter plot of branching fraction of $Z'$ to different decay channels. Right: scatter plot of $\sigma(pp\to Z')\times \text{BR}(Z'\to XY)$ at 13~TeV LHC. The solid line in the plot represents the ATLAS upper bound on 
$\sigma(pp\to Z')\times \text{BR}(Z'\to \ell^+\ell^-)$ at 95\% C.L.~\cite{Aaboud:2017buh}.} \label{fig:ZpBR}
\end{figure}
We note that $g_x' \lesssim 10^{-2}$ is sufficient to keep $\theta' < 10^{-3}$. This choice allows us to enhance the 
production of $Z'$ at a collider by four orders of magnitude, compared to the case when $g_x'=0$ where 
$\sin\theta' \sim 10^{-5}-10^{-6}$~(recall that $\sin\theta'$ depends on $g_x$ too). 
On the other hand, the coupling of $Z'$ with the heavy neutrinos is mainly governed by the choice of $g_x$. 
From Fig.~\ref{fig:thp} we can see that the value of $g_x$ can be taken to be $\mathcal{O}(0.1)$ while maintaining all 
relevant bounds.  If the mass of the heavy neutrino is less than $M_{Z'}/2$ then $Z'$ has an additional decay channel 
to a pair of heavy neutrinos. The decay to a pair of heavy neutrinos can be nearly 100\% while all other modes become 
significantly suppressed. In such a case the BR($Z'\to \ell^+\ell^-$) can be reduced to values less than 1\%. A scatter plot 
of the branching ratios of $Z'$ to different decay channels has been shown in the left panel of Fig.~\ref{fig:ZpBR}. 
Here we have varied $v_s$ between $1$--$10$~TeV while $g_x'$ is scanned over the range $0$--$0.02$.
On the right panel of the same figure, we show a scatter plot of $\sigma(pp\to Z')\times \text{BR}(Z'\to XY)$ at 13~TeV 
LHC. The solid line in the plot represents the ATLAS upper bound on the $\sigma(pp\to Z')\times \text{BR}(Z' \to \ell^+\ell^-)$ 
where $\ell = e, \, \mu$. As one can clearly see, this interplay actually helps us to produce $Z'$ at a higher rate while being within 
the bounds from the LHC in $Z' \to \ell^+ \,\ell^- $ mode~\cite{Aaboud:2017buh}. At the same time, we achieve a 
significantly high production cross-section of $NN$ through the $Z'$ resonance.

\begin{figure}
\centering
\includegraphics[width=0.5\textwidth]{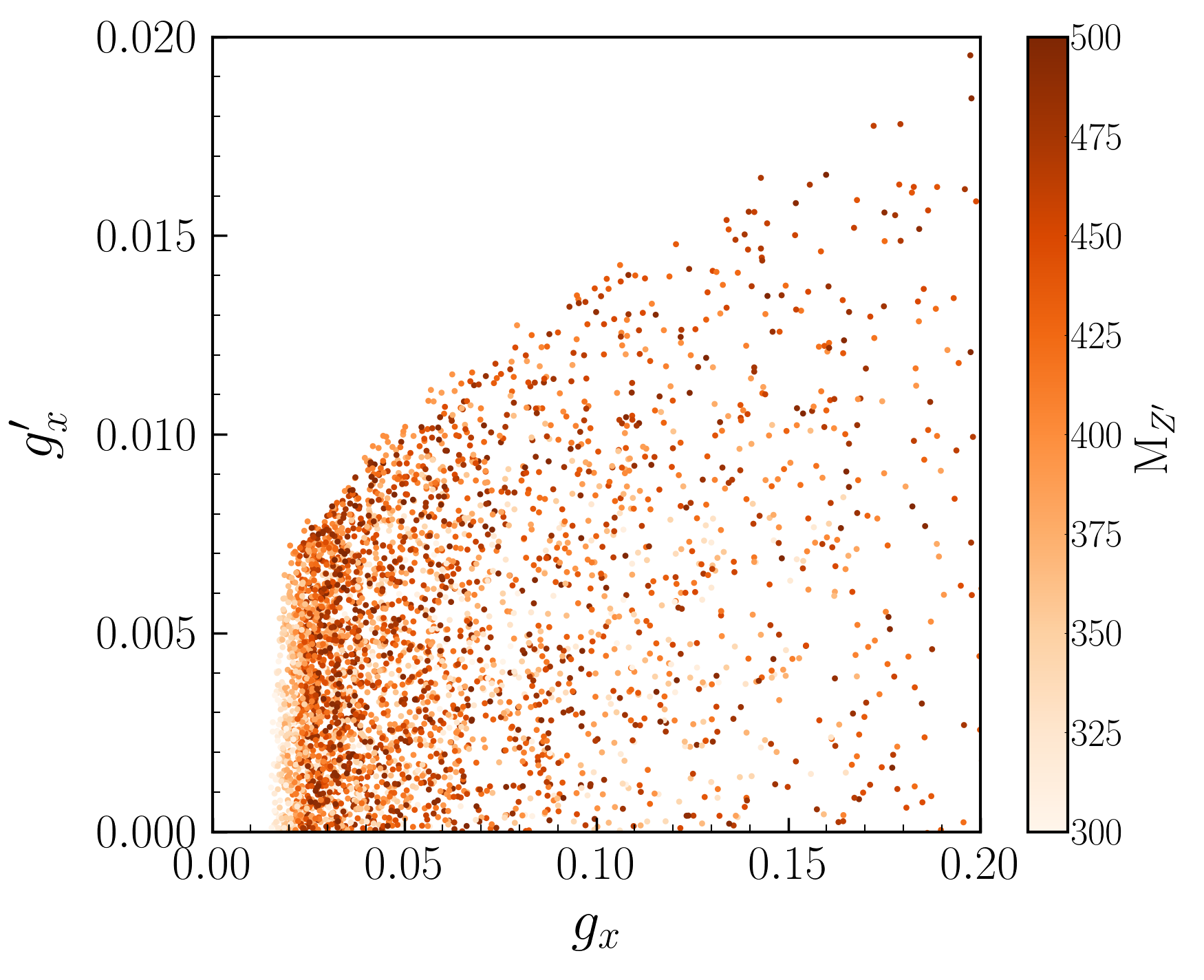}
\caption{Scatter plot of points satisfying all the three, {\it viz.} {\tt HiggsSignals}, {\tt HiggsBounds} and $Z'$ search in 
$\ell^+ \,\ell^-$ mode, in $g_x$-$g_x'$ plane. The color bar shows the variation of $M_{Z'}$.} \label{fig:bounds}
\end{figure}

In Fig.~\ref{fig:bounds}, we show a scatter plot of points which satisfy all the three, {\it viz.} {\tt HiggsSignals}, 
{\tt HiggsBounds} and $Z'$ search in $\ell^+ \,\ell^-$ mode, in $g_x$--$g_x'$ plane. The range for the scan over $v_s$ and $g_x'$ 
are the same as in Fig.~\ref{fig:ZpBR}. As expected, small $g_x$ and $g_x'$ values are always allowed as the 
constraint on $\theta'$ and constraint from $Z'$ searches are easily satisfied in that range of the 
parameter space. Since {\tt HiggsBounds} and {\tt HiggsSignals} limits do not have much dependence on $g_x$ and $g_x'$, 
they put little constraint in this plane. Higher values of $g_x'$ start getting disallowed since it leads to higher values 
for $\theta' > 10^{-3}$. However one finds that values of $g_x$ in the range of $0.01-0.2$ are allowed and 
$g_x' \lesssim 0.1 \, g_x$ is sufficient to suppress the $Z'$ decay to dilepton mode to avoid the constraints from the LHC,  
as can be seen in Figs.~\ref{fig:thp} and~\ref{fig:ZpBR}.   
\vspace{-8mm}
\section{Collider Analysis} \label{sec:collider}

We now look at the collider signatures for the new gauge boson $Z'$ at the LHC.  The most obvious signal for a heavy $Z'$ is 
via the Drell-Yan channel. In our scenario, the $Z'$ couples to the SM sector mostly through the mixing parameter and 
$g_x'$. Therefore, the on shell production rates of the $Z'$ are crucially dependent on the $\theta'$, which is also dependent 
on $g_x'$. For the gauge boson in the mass range of $200$--$500$~GeV, constraints indicate $\theta' \lesssim 10^{-3}$ which 
provides a significant limit to the production cross section of $\sigma (p \, p \to Z')$. However the cleanliness of the dilepton channel along with the resonant production of $Z'$ still provides a significantly 
strong constraint on $Z'$ mass.\footnote{The small $Z'$ width allows the use of {Narrow-Width 
Approximation~(NWA)} in calculating the di-lepton cross-section using $\sigma \times {\tt BR}$.} This bound can be relaxed if the $Z'$ decay to the 
charged lepton pair is suppressed, as shown in Fig.~\ref{fig:ZpBR}. The decay to a pair of heavy neutrinos opens up an 
interesting channel to search for $Z'$ in this model.  In addition we find that the upper bound on the production cross section 
$\sigma (p \, p \to Z')$ in this channel can be larger than what would be allowed in the absence of the 
$Z' \to NN$ decay.\footnote{Our choice of parameter space gives six heavy neutrinos ($\nu_k, \,\, k=4,5,\ldots ,8,9 $) of 
which four are taken to be heavier than $M_{Z'}$. The lighter ones are nearly degenerate in mass, which 
we identify as $ N ~ (\nu_4, \nu_5 \in N)$ in our analysis.}
Thus we focus on the $Z'$ signal through the pair production of heavy neutrinos via $Z'$ resonance~\cite{Huitu:2008gf,Basso:2008iv,FileviezPerez:2009hdc,Mansour:2012em,Khalil:2012gs,Abdelalim:2014cxa,Khalil:2015naa,Cox:2017eme,Accomando:2017qcs,Chiang:2019ajm}. Notably the pair production of heavy neutral leptons has also been looked at in the context of seesaw scenarios for neutrino mass~\cite{Aguilar-Saavedra:2009fxa,delAguila:2009bb,Kang:2015uoc,Das:2017deo} and some classes of $U(1)_X$ extensions with alternative 
charges to the more popular $U(1)_{B-L}$~\cite{Das:2017pvt,Das:2017flq}. 
The production of heavy Majorana neutrinos 
in the context of same-sign dilepton and multilepton searches have been carried out at LEP by DELPHI~\cite{DELPHI:1996qcc} 
and L3~\cite{L3:1992xaz,L3:2001zfe} Collaborations as well as at the LHC by CMS~\cite{CMS:2018iaf,CMS:2018jxx} and ATLAS 
Collaborations~\cite{ATLAS:2019kpx}. The searches look for heavy neutral lepton singly produced 
through the $Z$ boson at LEP and $W$ boson at the LHC, which then decays to a charged lepton and $W$. This mode translates 
into an upper bound on the mixing parameter $\big|V_{\ell\,N}\big|^2$ between the light neutrinos (flavor $\ell$) and the 
heavy neutrino. Note that in our case we can parametrize the off-diagonal $\big|V_{\ell\,N}\big|^2 \sim F^2$ as given 
in Eq.~(\ref{eq:vln}). As our $m_D \propto v_2$ and $\hat{M}_N \geq 100$ GeV, we have 
$F^2 \sim \big|V_{\ell\,N}\big|^2 \lesssim 10^{-6}$. This is much lower than the upper bound of $10^{-5} - 10^{-1}$ 
coming from the experimental data for $ 1~{\rm GeV} \, < \hat{M}_N < 1~{\rm TeV}$~\cite{CMS:2018jxx}, and allows us to 
choose heavy neutrino mass of $\mathcal{O}(100)$ GeV consistent with existing searches of heavy neutral leptons 
at experiments. In addition, pair production of neutral heavy leptons 
through a heavy resonance has also been studied at the LHC~\cite{CMS:2012axw,ATLAS:2019qrr}, where the heavy neutral leptons are long-lived giving rise to displaced vertex. These studies would however not constrain the parameter space as the 
heavy neutrinos have prompt decays in our study.  
Unlike the  other $U(1)$ extensions, 
$Z'$ in our case decays dominantly to a pair of heavy neutrinos while the $Z'$ production is driven by the close interplay 
of kinetic mixing between the two $U(1)$s and the $Z$-$Z'$ mixing arising out of symmetry breaking as the $Z'$ has no 
direct coupling with the SM quarks and charged leptons.

The dominant decay modes of $N$ are $\ell^{\pm}\, W^{\mp}$ and $\nu Z$. 
Since the heavy neutrinos are Majorana in nature, $N$ can decay to charged leptons with either sign. This gives an 
interesting set of possibilities for final states. Depending on the decay modes of $W^{\pm}$ and $Z$, we can have the following possibilities 
of final states.
\begin{itemize}
\item $4 \ell  + \slashed{E}_{T}$.
\item $3 \ell + 2 j +  \slashed{E}_{T}$.
\item $2 \ell + 4 j +  \slashed{E}_{T}$.
\item $4 j +  \slashed{E}_{T}$~(when only $N \to \nu Z$ decay is considered).
\end{itemize}
Although these are all interesting channels to look for $Z'$ in this model, especially the same-sign dilepton with jets and 
missing transverse energy~(MET), we mainly focus on the more sensitive four-lepton and three-lepton signals with smaller 
SM background in this article. Studies in the multilepton channels including the same-sign dilepton mode for heavy neutrinos produced 
via $Z'$ has always been of interest, and has been looked at 
before~\cite{Huitu:2008gf,delAguila:2008cj,Basso:2008iv,Perez:2009mu,Atre:2009rg,Accomando:2016sge,Cox:2017eme}. 

For our analysis of the trilepton and four-lepton channels, we have chosen three benchmark points. The values of the important 
parameters of these three benchmark points are tabulated in Table~\ref{tab:bps}. Note that the slight variation in the values of 
$v_s$ for the three benchmark points are made to adjust the $Z'$ mass to their respective values chosen for the analysis. 
The leading-order~(LO) production cross section of $Z'$ at the 14 TeV LHC run machine and branching ratios of $Z'\to NN$ for these three benchmark points are also mentioned in the table. Note that for $M_N > M_{Z'}/2$, the branching probability 
of BR$(Z' \to ee + \mu\mu) \sim 28\%$ constraining the allowed upper bound for $g_x'$ to become $2.48 \times 10^{-3}$, 
$4.58 \times 10^{-3}$, and $4.44 \times 10^{-3}$ for the three benchmark points, respectively. 
All these three points satisfy the constraints discussed in the last section. 

\begin{table}
\begin{center}\scalebox{1.0}{
\begin{tabular}{|c|c|c|c|}
\hline &  {\tt BP1} &  {\tt BP2} &  {\tt BP3}  \\ \hline
     $M_{Z'}$ (GeV) & 300      & 400        & 500           \\
    $M_{N}=\hat{M}_{N_{11}}$ (GeV) & 120      & 150        & 200           \\ \hline
     $g_x$                &  0.149  & 0.191     & 0.246        \\ 
$g_x ' \times10^{3}$ & 7.02     &  9.52      & 9.52           \\
$\tan\theta' \times10^{4}$  & 9.87    &  7.20   &  4.52      \\   \hline 
$\sigma(p \, p \rightarrow Z')$ (fb) &  215.5 &  148.2   & 67.7   \\  
BR $(Z' \rightarrow N \, N)$ &  0.987 & 0.985 & 0.990   \\
BR$(N \rightarrow \ell^\pm \, W^\mp (\nu Z) )$ & 0.75 (0.25) & 0.67 (0.29) & 0.60 (0.29) \\ 
\hline
 \end{tabular}}
\end{center}
\caption{Input parameters for the three benchmark points and the corresponding masses and mixing angles 
considered for our collider analysis (rounded off to the nearest digit). Note that we fix $\tan\beta =0.01$, $Y_{\nu_{11}}=0.05$, $Y_{\nu_{22}}=Y_{\nu_{33}}=0.2$, $Y_{L_{11}}=-10^{-9}$, $Y_{L_{22}}\simeq 5\times 10^{-8},  Y_{L_{33}}\simeq 2.8\times 10^{-7}$, and $\hat{M}_{N_{22}}=\hat{M}_{N_{33}}=1$~TeV for all 
benchmark points while $\lambda_s \simeq 0.884~(0.904)$ for {\tt BP1, BP2~(BP3)} and 
$v_s \simeq 1.01, 1.05, 1.02$ TeV for {\tt BP1, BP2, BP3}, respectively. }
\label{tab:bps}
\end{table}
Before discussing each specific analysis, we would like to mention the public packages that we have employed to 
perform the analysis.  The model was implemented in {\tt SARAH}~\cite{Staub:2013tta} to get the 
{\tt Universal Feynman Object~(UFO)}~\cite{Degrande:2011ua} files. {\tt SPheno}~\cite{Porod:2003um,Porod:2011nf} was 
used to generate the mass for the particle spectrum as well as the mixing parameters and mixing matrices connecting the 
gauge eigenstates to their mass eigenstates.  The UFO model files were then used to calculate the scattering process with 
{\tt Madgraph} and generate parton-level events with the {\tt MadEvent} event generator using the 
package {\tt MadGraph5@aMCNLO}~(v2.6.7)~\cite{Alwall:2011uj,Alwall:2014hca} at the LHC with 14~TeV center-of-mass energy. 
These parton-level events were then showered with the help of {\tt Pythia \!8}~\cite{Sjostrand:2014zea}. Detector effects were 
simulated using fast detector simulation in {\tt Delphes-3}~\cite{deFavereau:2013fsa} using the default ATLAS card. 
The final events were analyzed using the analysis package {\tt MadAnalysis5}~\cite{Conte:2012fm} to present our results.
\subsection{$ 4 \ell + \slashed{E}_T$ final state}
The $4 \ell$ final state is a relatively background free and clean event sample to study at the LHC. Some model dependent 
analysis has been carried out by experiments at the LHC to look for such final states~\cite{Aaboud:2018zeb,Aad:2021lzu}. 
We have checked that these analyses do not add any further constraints on our choice of the benchmark points. The 
four-lepton final state in our case occurs when both the $W$ and $Z$ bosons coming from each $N$, decay leptonically. 
In the case of $ N \to \ell \, W$ we expect MET from the neutrinos coming from the $W$ decay while the $N$ decays directly to 
neutrinos in the $Z$ channel.  Although the branching ratios of leptonic decay modes of $W$ and $Z$ is much smaller 
compared to their hadronic decay modes, higher charged lepton multiplicity in the final states are known to provide a 
cleaner signal with smaller SM background at a hadron collider. Thus the backgrounds for multilepton final states are 
manageable to negligible sizes at a hadron machine. This is one of the primary motivations behind the study of 
a  $4 \ell$ final state at the LHC. 

The major SM background for the $4 \ell + \slashed{E}_T$ final state comes from the following 
subprocesses~\cite{delAguila:2008cj}: 
\begin{align*}
p p \to V Z, && p p \to t\bar t Z, && p p \to VVV \,\,\, (V \equiv W^\pm, Z).
\end{align*}
All SM backgrounds were generated using the same event generator as in the 
case of the signal. We then scale the background cross section with their respective $k$ factors to make up for the 
next-to-next-to-leading-order (NNLO) corrections for $ZZ$ and NLO corrections for $t\bar tZ$ and $VVV$ backgrounds. The $k$ factors are taken to be 
$\simeq$ 1.72, 1.38, 2.01, and 2.27 for $ZZ$~\cite{Cascioli:2014yka},  $t\bar t Z$~\cite{Kardos:2011na},  
$WZ$~\cite{Grazzini:2016swo}, and $VVV$~\cite{Hong:2016aek,Yong-Bai:2016sal}, respectively.  

For our analysis, we choose events which have exactly $N_\ell = 4$ isolated charged leptons~($\ell = e, \mu$) in the final state.  
As basic acceptance cuts, we demand that all reconstructed objects are isolated~($\Delta R_{ab} > 0.4$). In addition,
\begin{itemize}
\item All charged leptons  must have $p_{T_\ell} > 10$~GeV and lie within the rapidity gap satisfying $|\eta_\ell| < 2.5$. 
\item We impose additional conditions to demand a hadronically quite environment by putting veto on events with light jets and 
$b$ jets  with $p_{T_{b/j}} > 30$~GeV and $|\eta_{b/j}| < 2.5$. This helps in suppressing a significant part of the 
background coming from $t \, \bar{t} (Z)$ production. 
\item We also demand a veto on any photon in the final state with $p_{T}^{\gamma} > 10$~GeV and $|\eta^{\gamma}| < 2.5$.
\end{itemize}
\begin{table}[h!]
\begin{center}\scalebox{1.0}{
\begin{tabular}{|c|c|c|c|}
\hline
Signal  & Cross section (fb) & SM Background  & Cross section (fb) \\ \hline \hline
 {\tt BP1}   &  0.688                &     $ZZ$                 &       9.088               \\       
 {\tt BP2}   &  0.476                &     $VVV$              &       0.111                \\
 {\tt BP3}   &  0.204                &     $W^{\pm} \, Z$  &       0.081               \\
                 &                           &     $t \bar{t} \, Z$    &       0.014               \\ \hline \hline
\end{tabular}}
\end{center}
\caption{The cross sections of signal and background for the final state $p p \rightarrow 4 \ell + \mET$ after 
the basic acceptance cuts and vetoes.}
\label{tab:4lcross}
\end{table}
We list the signal and background cross sections after the basic acceptance cuts on the charged leptons and the veto 
on additional light jets, $b$ jets and photons in the final state in Table \ref{tab:4lcross}. Note that with no requirement 
of MET in the final state, the dominant background comes from the $p \, p \to Z \, Z$ subprocess. 

\begin{figure}[h!]
\centering
\includegraphics[width=0.4\textwidth]{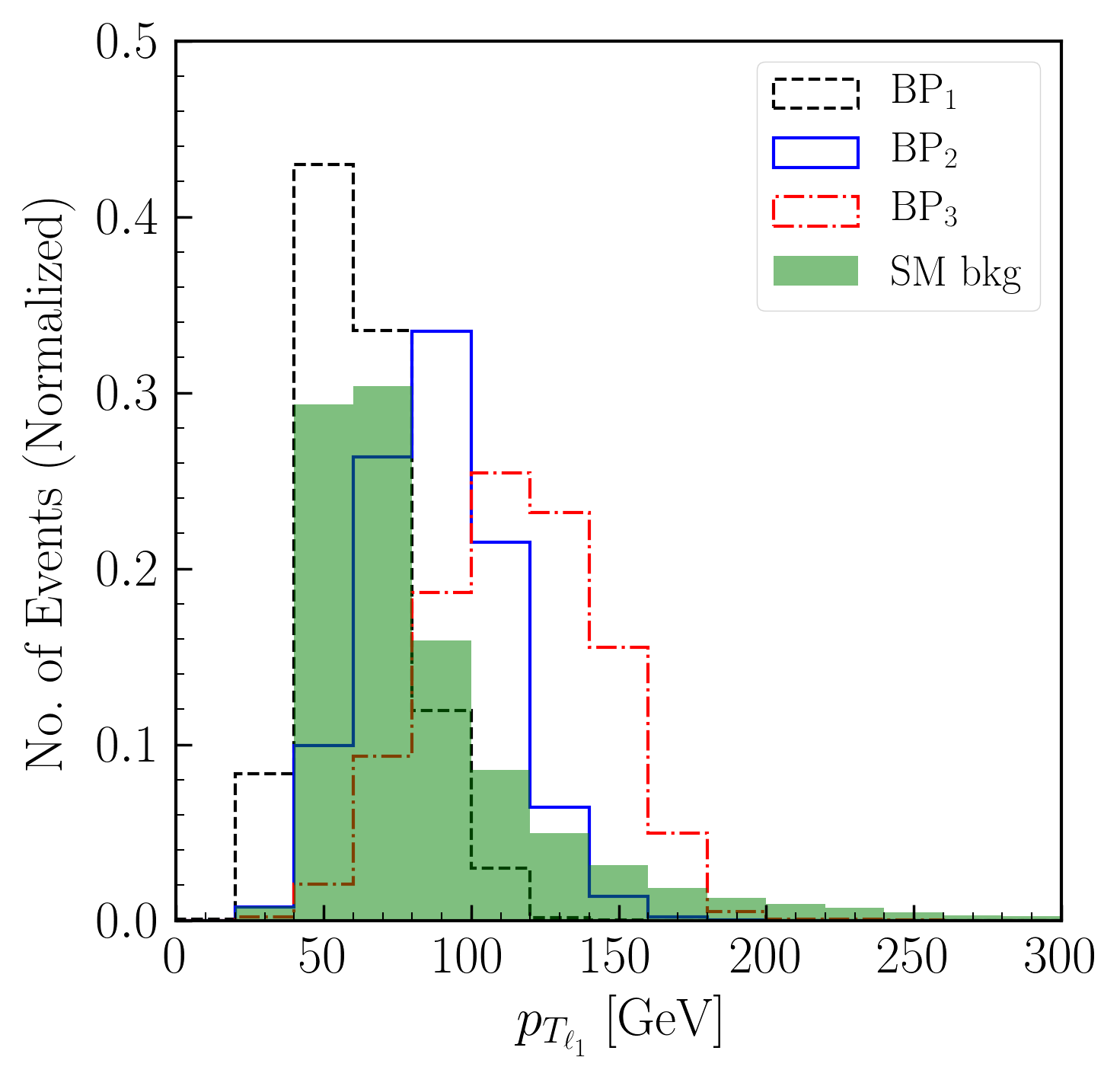} \hspace*{0.25in}
\includegraphics[width=0.4\textwidth]{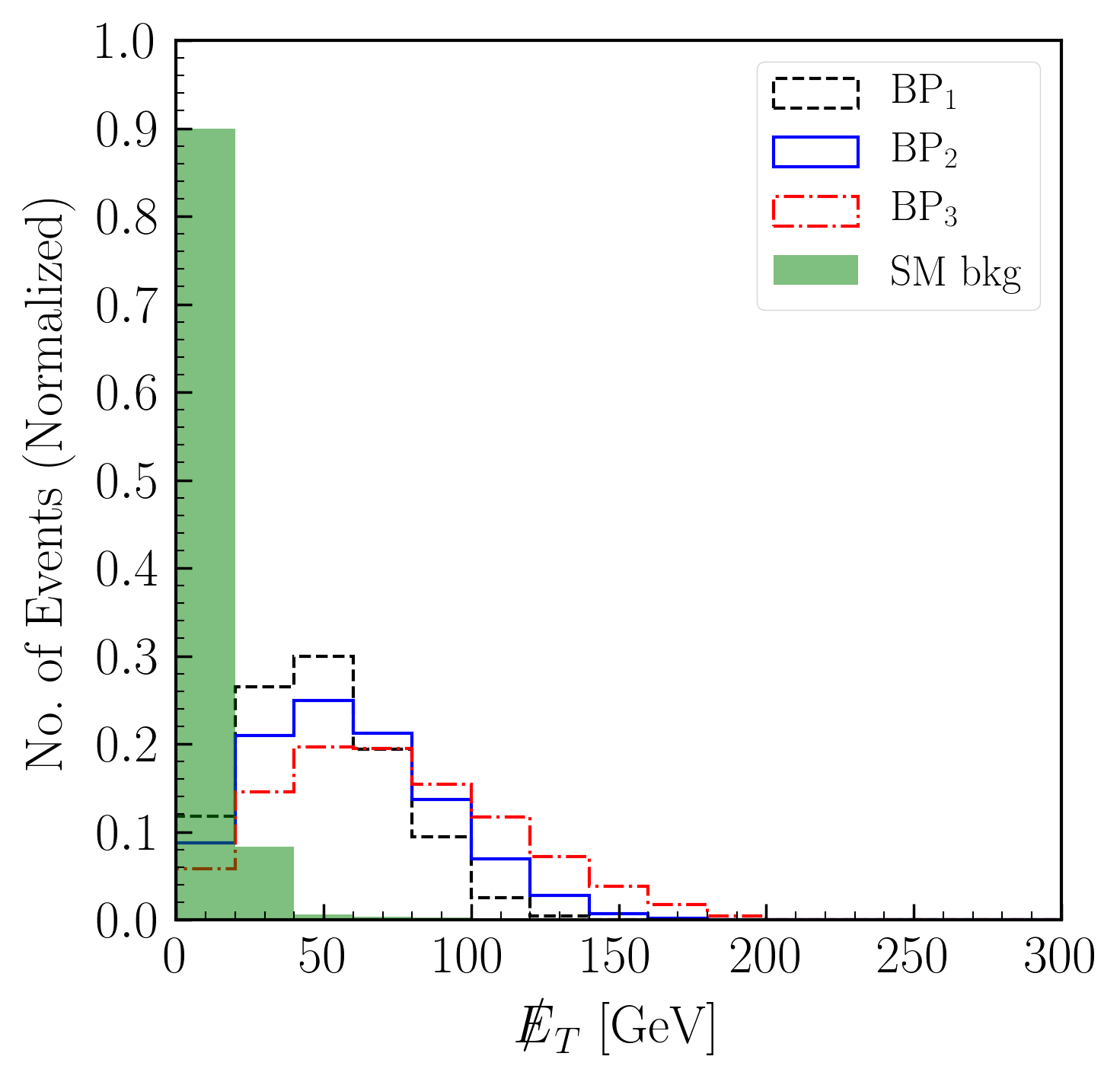}
\caption{Normalized distributions of $p_{T_{\ell_1}}$ (left panel) and $\mET$ (right panel) for $4 \ell + \mET$ final state 
at the 14 TeV LHC.} \label{fig:4lvarsA}
\end{figure}

\begin{figure}[h!]
\centering
\includegraphics[width=0.4\textwidth]{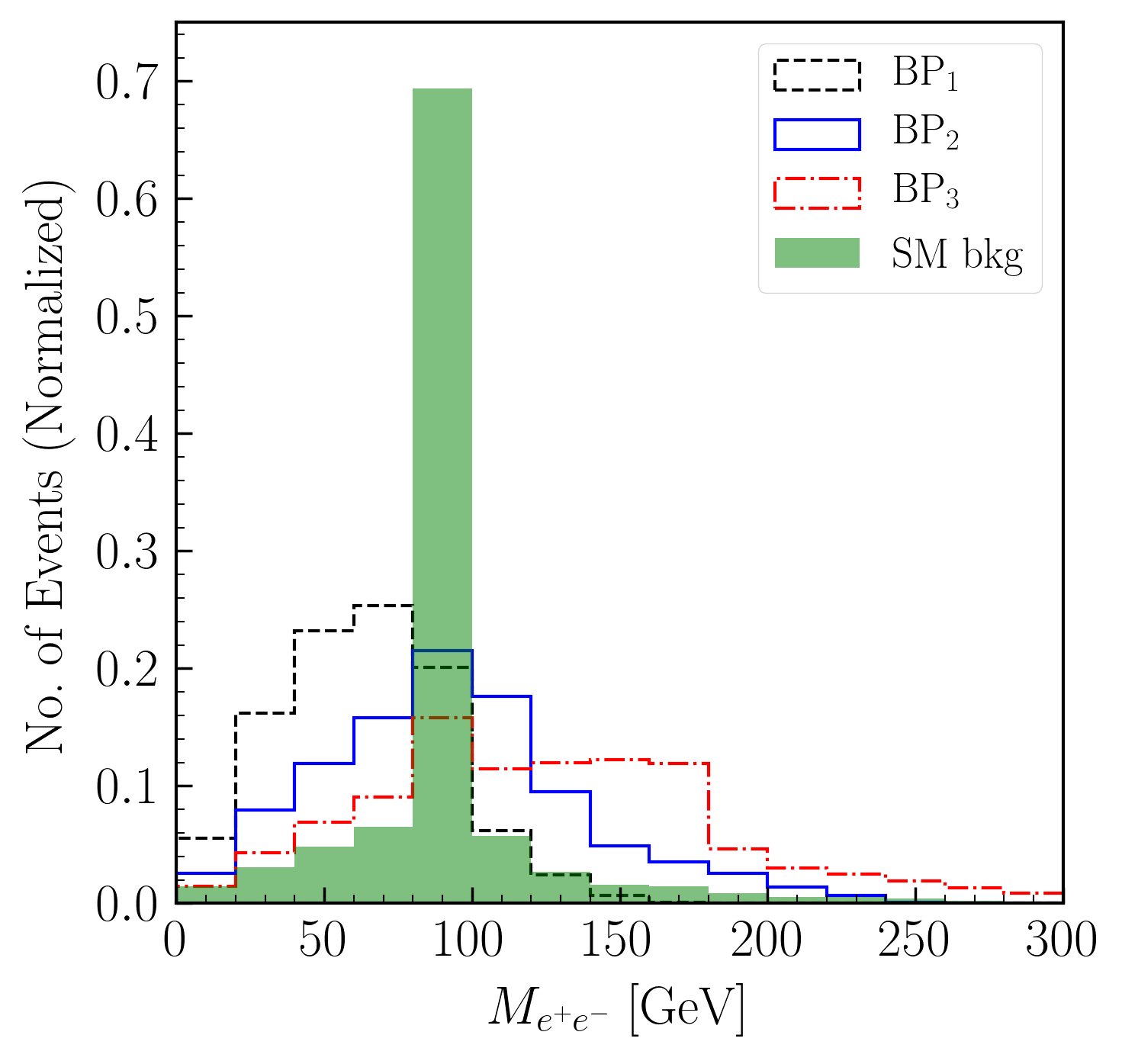} \hspace*{0.25in}
\includegraphics[width=0.4\textwidth]{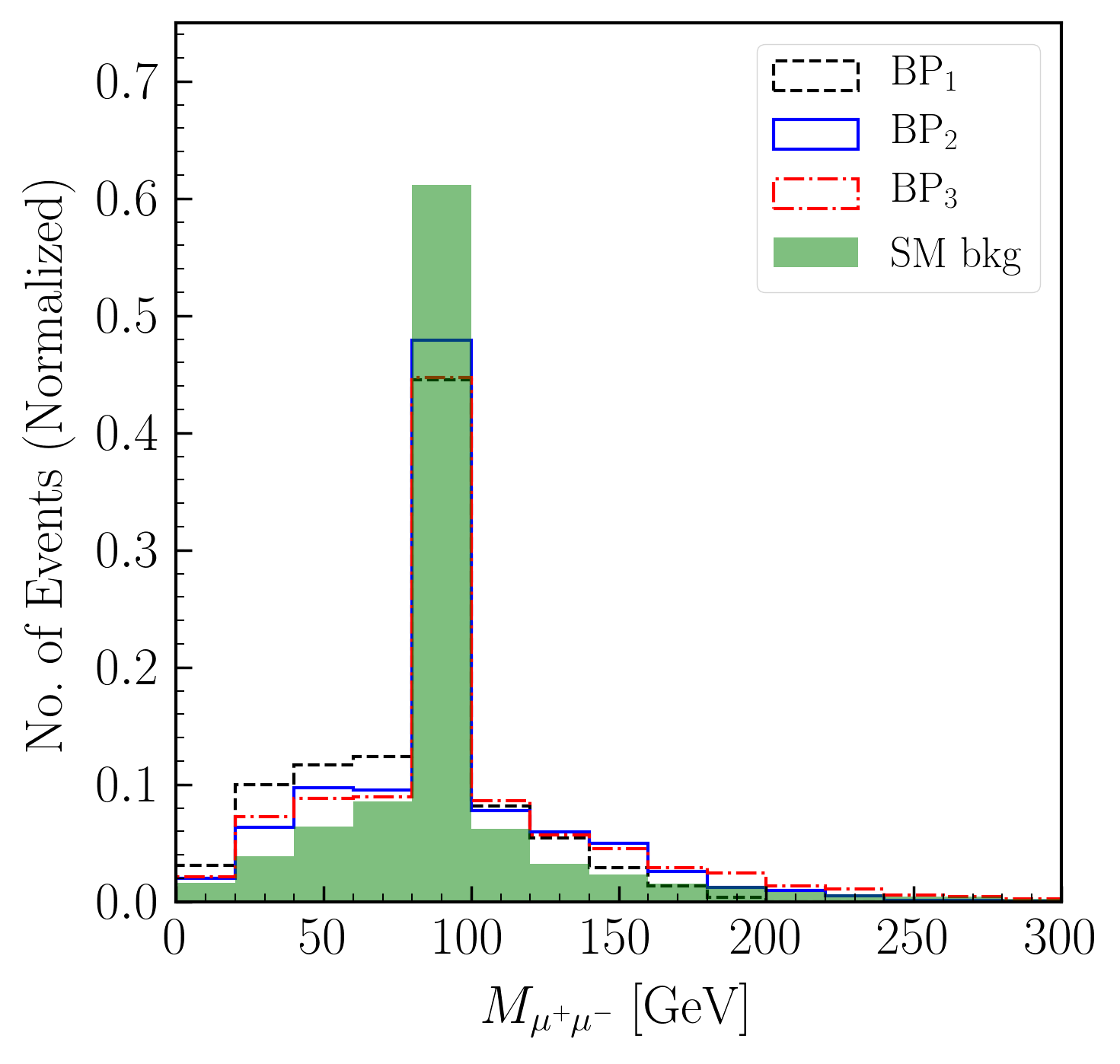}
\caption{Normalized distributions of $M_{e^+e^-}$ (left panel) and $M_{\mu^+\mu^-}$ (right panel) for $4 \ell + \mET$ final state 
at the 14 TeV LHC.} 
\label{fig:4lvarsB}
\end{figure}

To improve the signal to background ratio, one needs to exploit the kinematics of the signal events against that off the SM 
background. To achieve that, we must look at kinematic distributions of some relevant variables. In Figs.~\ref{fig:4lvarsA} and 
\ref{fig:4lvarsB}, we plot area normalized distributions for some of these important kinematic variables after detector 
simulation. In the left panel of Fig.~\ref{fig:4lvarsA},  we note that the $p_T$ distribution of the leading charged lepton peaks 
around $40$--$50$~GeV for {\tt BP1},  around $80$--$90$~GeV for {\tt BP2} and around $100$--$120$~GeV for {\tt BP3}. 
These peaks are 
consistent with the mass difference between $N$ and $W$~($M_N - M_W$) for the three benchmark points~(BPs) implying 
that the leading lepton
comes from the primary decay of the heavy neutrino. We also note that with higher mass difference one expects to get the 
peak at a higher value of $p_T$ for the signal. Thus a stronger $p_T$ cut on the leading lepton would help remove the 
SM backgrounds with leading leptons on the softer side compared to the signal. However, the charged leptons in the SM 
background originate from the $Z$ and $W$ bosons and also show a peak around $p_T \sim M_{Z/W}/2$ leading to a significant 
overlap with that of the signal events of {\tt BP1} and to some extent with that of the remaining two BPs too. The overlaps are 
significantly larger for the subleading leptons. Thus we choose a moderately smaller $p_T > 20$~GeV requirement on the 
leading lepton, while all the remaining three leptons have $p_T > 10$~GeV.
The other important distributions correspond to the MET~($\slashed{E}_T$) distribution and the invariant mass of the pair of oppositely charged same flavor~(OSSF) leptons {\it viz.} $M_{e^+e^-}$ and $M_{\mu^+\mu^-}$.
Note that for the $ZZ$ background, the only source of MET would come from the imbalance in the visible $p_T$ arising 
out of the mismeasurement of jet and lepton energies.  Thus a MET cut of $\slashed{E}_T > 15$~GeV helps us remove the 
$ZZ$ background to a great extent without affecting the signal too much. The plot in the right panel of 
Fig.~\ref{fig:4lvarsA} supports this expectation. Note that as the particle spectrum is light and the corresponding decay products
do not carry too much $p_T$ we put an upper bound of 200~GeV on the $p_T$ of the leading lepton and $\mET$ which helps in 
suppressing some SM background. The effect of the aforementioned selection cuts are shown in Table \ref{tab:cutflow4l}. 

The invariant mass of $e^+e^-$ and $\mu^+\mu^-$ are shown in Fig.~\ref{fig:4lvarsB}. We note that the signal events 
would not show a peak around the $Z$ boson mass unless $N$ decays via the~($\nu \, Z$) mode. For the backgrounds, 
the invariant mass of OSSF  leptons peak at the $Z$ boson mass. A large fraction of the signal events comes 
from $N \to e \, W$ decay mode. Thus an invariant mass cut on the OSSF leptons of electron type should be more 
useful in removing that background. However, as the  $p_{T_{e^\pm}}$ of the signal events are not very hard, we observe 
an overlap of the $Z$ peak with the signal events in the $M_{e^+e^-}$ distribution. So a cut of $Z$ peak in the $e^+e^-$ 
mode does not help a lot in improving the signal to background ratio. On the other hand, we expect that the fraction of 
events for the signal that contain at least a $\mu^+\mu^-$ pair will be much smaller~[$\sim (28-31)\%$ for the 3 BPs] 
when compared to the full $4\ell$ mode~(as evident from the branching fractions of $N$ and $Z$). In contrast, the 
background is expected to be equally divided in the $e$ and $\mu$ modes. So although the normalized distribution in 
$M_{\mu^+\mu^-}$ distribution shows a significant part of the signal in the mass bin of $Z$ peak, we must realize that 
the distribution only corresponds to a very small fraction of the $4\ell + \slashed{E}_T$ events after cuts. Therefore a 
cut to remove the $Z$ peak in the $\mu^+\mu^-$ distribution 
($ 80 < M_{\mu^+\mu^-} < 95$~GeV)\footnote{The reason for an asymmetric cut around the $Z$ mass is based on the fact 
that the invariant mass distribution from a resonant production always falls more rapidly beyond the parent particle 
mass.} helps in suppressing a significant part of the SM background and improves the signal significance. To facilitate 
this we also demand that the four-lepton final state signal has at most a single pair of $\mu^+\mu^-$. 

The result of the analysis and the respective selection cuts are presented in Table \ref{tab:cutflow4l} for an integrated
luminosity of $\mathcal{L} = 100$~fb$^{-1}$ at the 14 TeV LHC.    
%
%
\begin{table}[t!]
	\centering
		\resizebox{15cm}{!}{
	\begin{tabular}{|c|c|c|c|c|c|c|c|}
	\cline{2-8}
	\multicolumn{1}{c|}{$\mathcal{L}=100$ fb$^{-1}$} 
      & \multicolumn{4}{c|}{SM-background} &  \multicolumn{3}{c|}{Signal } \\  \hline	
                         Cuts  & ~~$ZZ$~~ & ~~$VVV$~~   & ~~$t\bar t Z$~~ &  ~~$W^{\pm}Z$~~ 
                                  &  ~~{\tt BP1}~~ & ~~{\tt BP2}~~ & ~~{\tt BP3}~~ \\ \hline		
     $N_{\mu} \leq 2$   & 566.5 & 5.69 &  0.53  &  4.52 & 64.5  & 43.7 &18.7 \\ \hline	
     $(15 < \mET < 200)$~GeV & 107.3 & 4.8   &  0.47  &  3.97 & 60.07  & 41.66 & 18.04 \\  \hline	         		 
     $(20 < p_{T_{\ell_1}} < 200)$~GeV    & 103.7 & 4.19 & 0.38   & 3.97 & 60.01  & 41.66 &18.02 \\  \hline	        
     $M_{\mu^+\mu^-} < 80$~GeV or $M_{\mu^+\mu^-} >95$~GeV  &35.35 & 2.74 & 0.25 & 3.6 & 56.17  & 38.5  &16.6 \\  \hline	
   \multicolumn{1}{|c|}{Total Events after cuts} 
      & \multicolumn{4}{c|}{41.94} & 56.17  & 38.5  &16.6 \\  \cline{1-8}  \hline	         
      & \multicolumn{4}{c|}{Significance ($\mathcal{S}$)}  
      & 7.38 & 5.67  & 2.42 \\ \hline \hline
	\end{tabular}}
	\caption{ The cut-flow information on the $p \, p \rightarrow 4 \ell+\slashed{E}_T$ process for both the signal and 
	background along with the significances for {\tt BP1, BP2, BP3} at the 14 TeV LHC for 100 fb$^{-1}$ integrated luminosity.}
	\label{tab:cutflow4l}
\end{table}	

We calculate the signal significance~($\mathcal{S}$) by using the following formula.
\begin{equation}
\mathcal{S} = \sqrt{2\left[\left(S+B\right)\ln\left(\frac{S+B}{B}\right)-S\right]},
\end{equation}
where $S$ and $B$ are number of signal and background events, respectively. The signal significance for these three benchmark points are provided in the last column of Table~\ref{tab:cutflow4l}. 
We can see that the signal for {\tt BP1} and {\tt BP2} have quite significant discovery potential as they correspond to a 
lighter $Z'$ compared to {\tt BP3}. The signal significance for a lighter $Z'$ is high even with 50~fb$^{-1}$ integrated luminosity, 
which may however be constrained by the current LHC data. On the other hand such a constraint may be avoided by slight 
modification of the $Z$-$Z'$ mixing, as in the case of the dilepton Drell-Yan channel. The important aspect of the above analysis 
however lies in the fact that signals for a light $Z'$, which does not talk to the SM particles directly may be absent in the 
dilepton or dijet modes but can be discovered in a more exotic $4 \ell+\slashed{E}_T$ channel.

\subsection{$3\ell + 2j + \mET$ final state}
We now focus on the final state with a larger production rate as compared to the $4\ell$ final state, {viz.} the 
$3\ell + 2j + \mET$ signal at the LHC~\cite{CMS:2018iaf,CMS:2018jxx,ATLAS:2019kpx}. However this channel has 
little advantage over the $4\ell$ mode since the background events also become larger in this channel. The main SM 
background comes from the following subprocesses~\cite{delAguila:2008cj}: 
\begin{align*}
p p \to V Z, && p p \to t\bar t + t\bar t \, Z , && p p \to VVV \,\,\, (V \equiv W^\pm, Z).
\end{align*}
As before, we include $k$ factors for the LO cross section for the SM background to account for the NNLO correction for 
$WZ$ and $t\bar t$ and the NLO correction for $VVV$ and $t\bar t Z$ backgrounds. 
The $k$ factor is $\simeq 1.6$ for $t\bar t$~\cite{Czakon:2013goa}.  

The object reconstruction to identify the final state particles is similar to what was done for the $4 \ell + \mET$ final state. The 
basic acceptance cuts considered for the $3\ell + 2j + \mET$ signal are that all reconstructed objects are isolated 
($\Delta R_{ab} > 0.4$) and satisfy the following requirements.
\begin{itemize}
\item We have exactly three charged leptons, $N_{\ell}=3$ ~($\ell = e, \mu$) in the final state, each with $p_{T_\ell} > 10$~GeV 
and lying within the rapidity gap $|\eta_\ell| < 2.5$. 
\item We have exactly two light jets, $N_j= 2$  in the final state, each with $p_{T_j} > 30$~GeV 
and lying within the rapidity gap $|\eta_j| < 2.5$.
\item We impose veto on events with a $b$ jet having $p_{T_{b}} > 30$~GeV and $|\eta_{b}| < 2.5$. This again helps in 
suppressing a significant part of the background coming from $t \, \bar{t} (Z)$ production. 
\item We also demand a veto on any photon in the final state with $p_{T}^{\gamma} > 10$~GeV and 
$|\eta^{\gamma}| < 2.5$.
\end{itemize}

\begin{table}[t!]
\begin{center}\scalebox{1.0}{
\begin{tabular}{|c|c|c|c|}
\hline
Signal  & Cross section (fb) & SM Background  & Cross section (fb) \\ \hline \hline
 {\tt BP1}   &  1.723                &     $ZZ$                 &       1.528               \\       
 {\tt BP2}   &  1.526                &     $VVV$              &       0.266                \\
 {\tt BP3}   &  0.717                &     $W^{\pm} \, Z$  &       37.23                \\
                 &            &     $t \bar{t} + t \bar{t} \, Z$   &       1.745               \\
\hline \hline
\end{tabular}}
\end{center}
\caption{The cross sections of signal and background for the final state $p p \rightarrow 3 \ell + 2j + \mET$ after 
the basic acceptance cuts and vetoes.}
\label{tab:3lcross}
\end{table}
We list the signal and background cross sections after the basic acceptance cuts on the charged leptons, jets, and  
a veto on any $b$ jet and photons in the final state in Table \ref{tab:3lcross}. We find that with the $b$-jet veto the 
$t\bar t$ cross section becomes quite small whereas the leading background comes from the $WZ+\,$jets final state 
where both the gauge bosons decay leptonically to give three charged leptons in the final state. For the signal, we again
expect the dominant contribution to come from the $N \to e \, W$ decay mode, where one of the $W$ decays hadronically to 
two jets.

\begin{figure}[t!]
\centering
\includegraphics[width=0.407\textwidth]{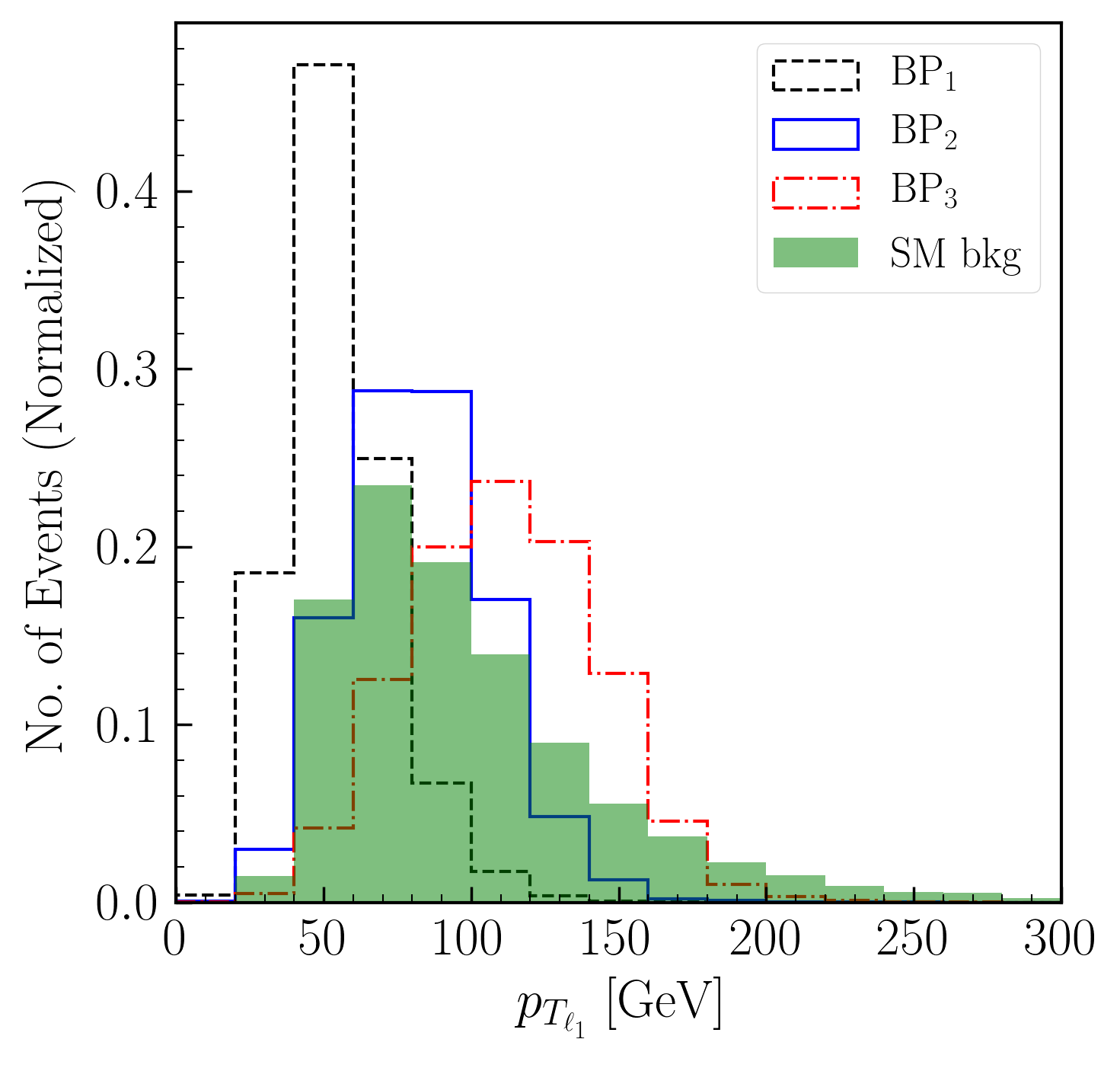}  \hspace*{0.25in}
\includegraphics[width=0.4\textwidth]{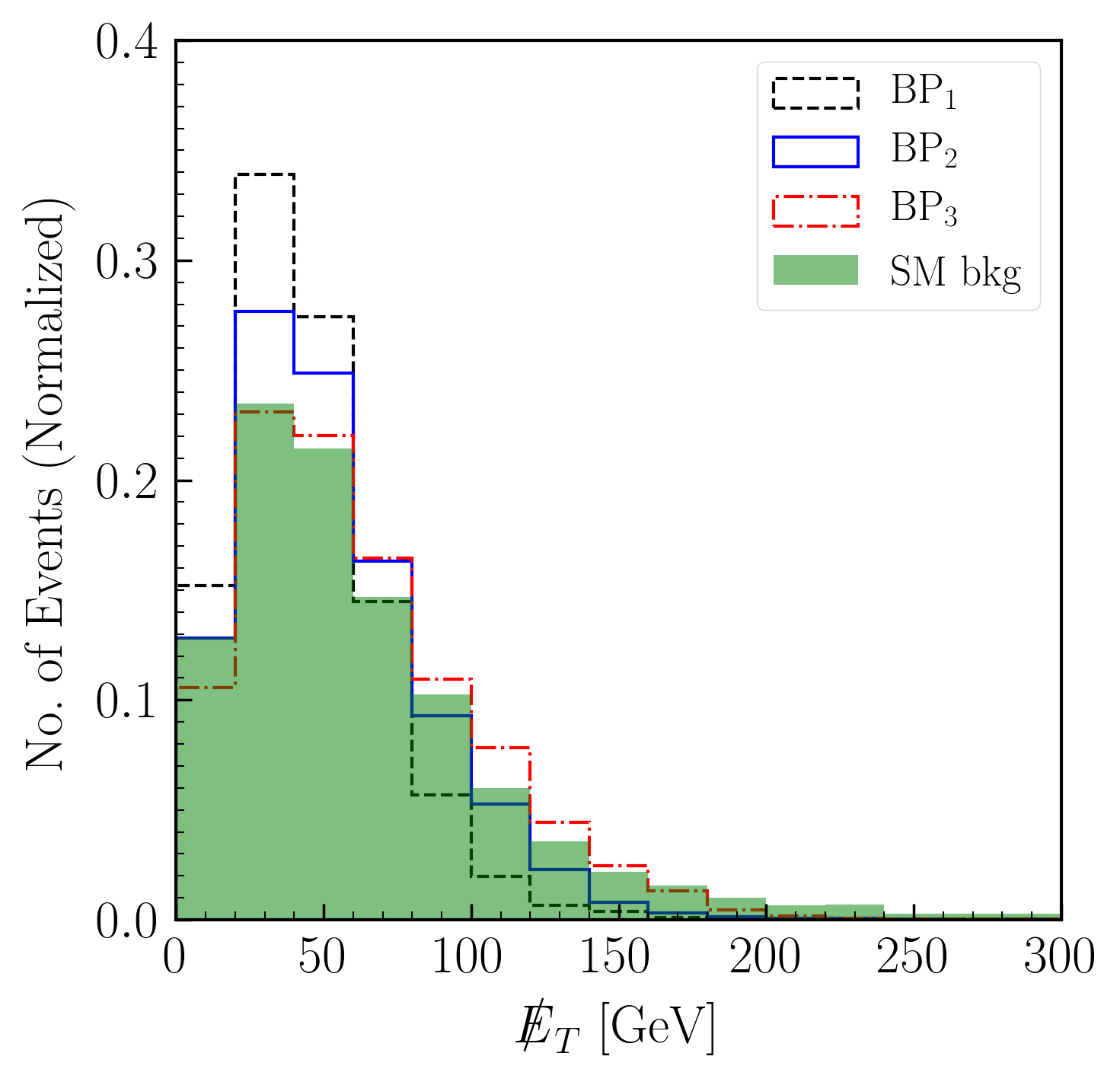}
\caption{Normalized distributions of $p_{T_{\ell_1}}$ (left panel) and $\mET$ (right panel) for $3 \ell + 2j + \mET$ final 
state at the 14 TeV LHC.} 
\label{fig:3lvarsA}
\end{figure}
\begin{figure}[t!]
\centering
\includegraphics[width=0.4\textwidth]{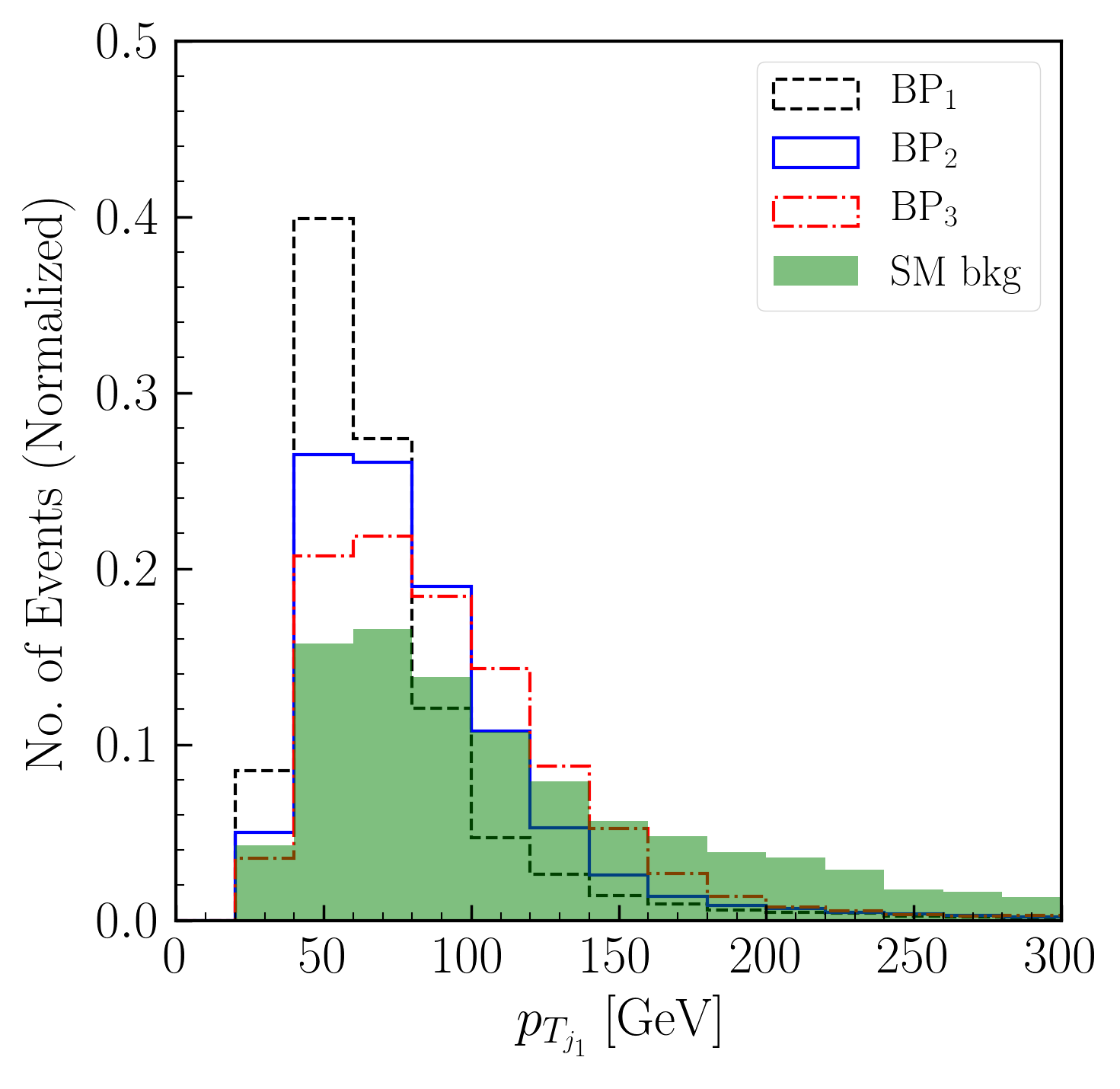}  \hspace*{0.25in}
\includegraphics[width=0.407\textwidth]{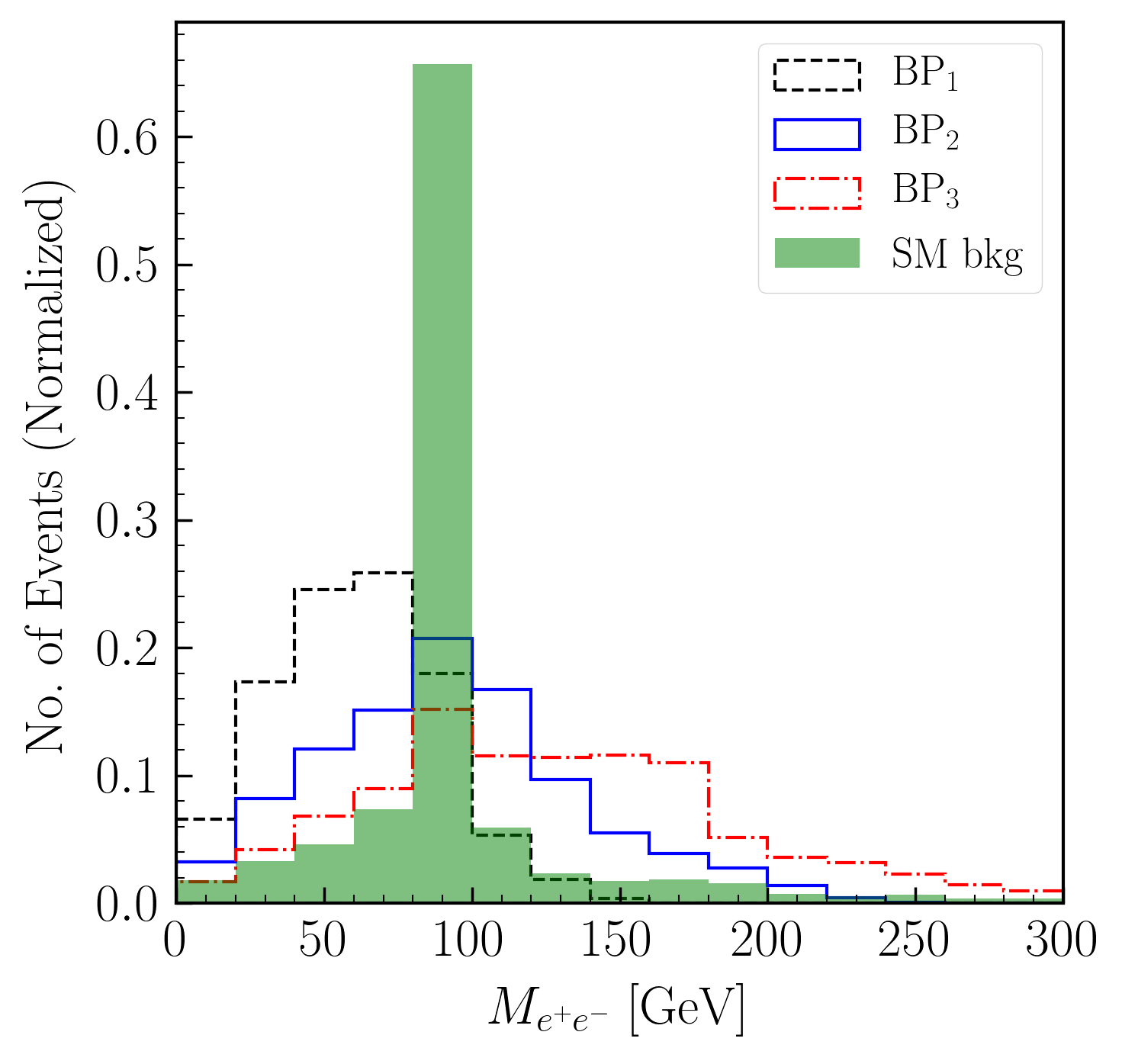}
\caption{Normalized distributions of $p_{T_{j_1}}$ (left panel) and $M_{e^+e^-}$ (right panel) for $3 \ell + 2j + \mET$ final 
state at the 14 TeV LHC.} 
\label{fig:3lvarsB}
\end{figure}

As we note that the signal is rich in $e^\pm$ and the $\mu$ multiplicity peaks at one, it again seems beneficial to 
put a constraint on $N_\mu \leq 1$ which should not affect the signal too much while suppressing the SM background. 
This can be seen from the cut-flow numbers presented in Table \ref{tab:cutflow3l}. We now look at the 
distributions of some of the important variables for this final state which are shown in Figs.~\ref{fig:3lvarsA} and \ref{fig:3lvarsB}. 
In Fig.~\ref{fig:3lvarsA} we plot the $p_T$ distribution of the leading lepton as well as the $\mET$ distribution. The lepton $p_T$
shows a similar behavior to the case of $4\ell$ final state and therefore we stick to a similar selection cut on the leading lepton 
to have $p_T > 20$~GeV. The $\mET$ distribution is markedly different due to the contributions from other background processes
dominating over the ones that contributed to the $4\ell + \mET$ case. However, we still note that the $\mET > 15$~GeV cut will 
suppress the $ZZ$ background as seen in Table \ref{tab:cutflow3l}. As in the case of $4\ell+\mET$ we again put an upper bound 
of 200~GeV on the $p_T$ of the leading lepton and $\mET$ to suppress the SM background which has a longer tail 
in the distributions extending beyond 200~GeV.

In Fig.~\ref{fig:3lvarsB} we plot the $p_T$ of the leading jet and the invariant mass distribution in $e^+e^-$. As the jets 
for the signal are not expected to be hard, we put an upper bound on them as $p_{T_{j_1}} <  200$~GeV. The dominant 
suppression in the background comes from the invariant mass cut where we remove the $Z$ peak. As we expect the 
electron or positron~($e$) to come from the decay of $N$ for the signal, we expect no $Z$ peak in the signal. 
Thus the invariant mass cut along with the constraint on $\mu$ multiplicity proves to be the most important condition 
that improve the $S/B$ for the $3 \ell + 2j + \mET$ final state.

The result of the analysis and the respective selection cuts are presented in Table \ref{tab:cutflow3l} for an integrated
luminosity of $\mathcal{L} = 100$ fb$^{-1}$ at the 14 TeV LHC. We can see that, as in the case of $4\ell + \mET$, the signal 
for {\tt BP1} and {\tt BP2} again has quite large significance, albeit slightly smaller for the same integrated luminosity.  
The above analysis however shows that both the $3\ell$ and $4\ell$ final states show a promising discovery channel 
for light $Z'$ which does couple to the SM particles directly, with the higher lepton multiplicity case doing slightly better. 
\begin{table}[t!]
	\centering
		\resizebox{15cm}{!}{
	\begin{tabular}{|c|c|c|c|c|c|c|c|}
		\cline{2-8}
		\multicolumn{1}{c|}{$\mathcal{L}=100$ fb$^{-1}$} 
      & \multicolumn{4}{c|}{SM-background} &  \multicolumn{3}{c|}{Signal } \\  \hline	
                         Cuts  & ~~$W^{\pm}Z$~~ & ~$t\bar t + t\bar t Z$~  & ~~$Z Z$~~ 
                                  &  ~~$VVV$~~ &  ~~{\tt BP1}~~ & ~~{\tt BP2}~~ & ~~{\tt BP3}~~ \\ \hline		
     $N_{\mu} \leq 1$   & 2246.0 & 147.2 & 86.5  & 26.0 & 170.4  & 150.6 &70.7 \\ \hline	
     $(15 < \mET < 200)$~GeV & 2022.0 & 146.2   &  39.0  &  22.1 & 155.0  & 139.4 & 66.1 \\  \hline	  
     $ p_T^{j_1} < 200$~GeV   & 1686.0 & 119.3 & 35.7   & 18.8 & 152.1  & 135.8 & 64.0 \\  \hline	           		 
     $(20 < p_{T_{\ell_1}} < 200)$~GeV   & 1608.0 & 118.7 & 34.6  & 17.2 & 151.4  & 135.7 & 63.7 \\  \hline	        
     $M_{e^+e^-} < 85$~GeV or $M_{e^+e^-} >95$~GeV  & 228.0 & 97.3 & 4.9 & 2.2 & 124.9  & 96.0  & 49.0 \\  \hline	
       \multicolumn{1}{|c|}{Total Events after cuts} & \multicolumn{4}{c|}{332.4} & 124.9  & 96.0  & 49.0 \\  \cline{1-8}  \hline	         
      & \multicolumn{4}{c|}{Significance ($\mathcal{S}$)}  
      & 6.48 & 5.04  & 2.63 \\ \hline 
	\end{tabular}}
	\caption{ The cut-flow information on the $p \, p \rightarrow 3 \ell+ 2j + \slashed{E}_T$ process for both the signal and 
	background along with the significances for {\tt BP1, BP2, BP3} at the 14 TeV LHC for 100 fb$^{-1}$ integrated luminosity.}
	\label{tab:cutflow3l}
\end{table}	
The analysis can be extended to include heavier $Z'$ as well and consider the other final states available for the 
$Z'$, which would be similar to the more traditional $Z'$ searches such as the $U(1)_{B-L}$ models for 
 example~\cite{Huitu:2008gf,Basso:2008iv}.
\section{Summary and Outlook} \label{sec:summary}

We consider a neutrinophilic model as an extension of the SM by introducing a $U(1)$ group 
which couples directly to only heavy neutral fermions, singlet under the SM. The neutral fermion charged under the 
new group couples to the SM matter fields through Yukawa interactions via a neutrinophilic scalar doublet. The neutrinos 
in the model get their mass from a standard inverse-seesaw mechanism while an added scalar sector is responsible 
for the breaking of the gauged $U(1)$ leading to light neutral gauge boson~($Z'$). We study the phenomenology of 
having such a light $Z'$ in the context of neutrinophilic interactions as well as the role of allowing kinetic mixing 
between the new $U(1)$ group with the SM hypercharge group. We show that current experimental searches 
allow for a very light $Z'$ if it does not couple to SM fields directly and highlight the search strategies at the LHC. 

To highlight the features of the model, we calculate the mass and mixing of the scalar, gauge and matter fields 
after symmetry breaking and look at the experimental constraints on the model parameters. We find that once
the scalar sector is set to agree with the Higgs searches, by choosing the lightest $CP$-even scalar to be the 125~GeV
SM Higgs boson, the $Z'$ phenomenology is only dependent on the $Z$-$Z'$ mixing and its coupling to the heavy neutral 
fermions. Following an examination of the allowed region for the mixing angle and the $U(1)_X$ gauge coupling we 
determine two regions of parameter space depending upon the value of $\tan\beta$, the ratio of the doublet VEVs. 
For $\tan\beta > 1$ we find an upper bound on the ratio $v_2/v_1 < 3$ from the perturbativity requirement on the
fermion-fermion scalar couplings. We also observe that $g_x$ and $g_x'$ are of the same order when $\tan\beta > 1$,
which gives us a $Z'$ phenomenology driven by the $Z$-$Z'$ mixing angle $\sin\theta'$ with the dominant decay to SM
fermion pair. A more interesting scenario emerges for $\tan\beta < 1$ where the $g_x'$ and $g_x$ are no longer required 
to be of the same order anymore.  We find that the $Z'$ signatures are now dependent on the interplay of the 
$Z$-$Z'$ mixing as well as the  $U(1)_X$ gauge coupling $g_x$ which is allowed to be large. Thus the $Z'$ can now decay 
dominantly to a pair of heavy neutrinos while the $Z'$ is produced through the $Z$-$Z'$ mixing parameter driven by $g_x'$.
We analyze the signal for such a scenario at the LHC with $\sqrt{s}=14$ TeV in the $4\ell + \mET$ and $3\ell + 2j +\mET$
channels for a $Z'$ lying in the mass range $200$--$500$~GeV. We find that although the dilepton Drell-Yan channel is 
much suppressed here, the discovery prospects of observing a neutrinophilic $Z'$ is significantly high in the above channels.
We show the significance of the signal using an integrated luminosity of 100 fb$^{-1}$ for three benchmark points.
We conclude that multilepton final states could be crucial in discovering such a neutrinophilic gauge boson lying in the 
mass range of $200$--$500$~GeV with even a very tiny gauge-kinetic mixing of the order $\mathcal{O}(10^{-3})$.

We must point out here that other interesting signatures of the $Z'$ in such a model is being left for future work, which 
include flavor violating decays of the $Z'$, a more detailed analysis of the scalar sector with the $Z'$ and implications 
of a very light $Z'$, and a singlet scalar~\cite{Abdallah:2021dul}.

\section*{Appendix: Coupling of $Z'$ gauge boson with fermions}

Below, we list the coupling of the $Z'$ gauge boson with the fermions in the model. We define $s_W \equiv \sin\theta_W$ 
and $c_W \equiv \cos\theta_W$ where $\theta_W$ is the Weinberg angle while  $s_{\theta'} \equiv \sin\theta'$ and
$c_{\theta'} \equiv \cos\theta'$ where $\theta'$ is the $Z$-$Z'$ mixing angle. In addition, $T_3$ and $Q_f$ represent the isospin and electric charge of the fermions, respectively, while $P_{L/R} = \frac{1 \mp \gamma_5}{2}$ are the projection 
operators.
\begin{figure}[h!]
\includegraphics[width=0.19\textwidth]{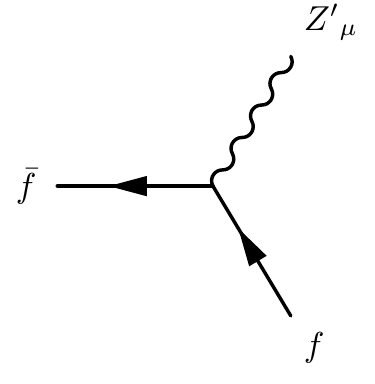}
\end{figure} 
\small{
\begin{align*} 
 i\left(\frac{e \, s_{\theta'}}{s_{W} c_{W}} \left(T^3 - Q_f s^2_{W}\right) + g_x' c_{\theta'}\left(T^3 - Q_f\right)\right)\!\gamma^\mu P_L - i\left(\frac{e \, s_{\theta'}}{s_{W} c_{W}} Q_f s^2_{W} + g_x' c_{\theta'}Q_f\right)\!\gamma^\mu P_R 
 \end{align*}  }

\begin{figure}[h!]
\includegraphics[width=0.19\textwidth]{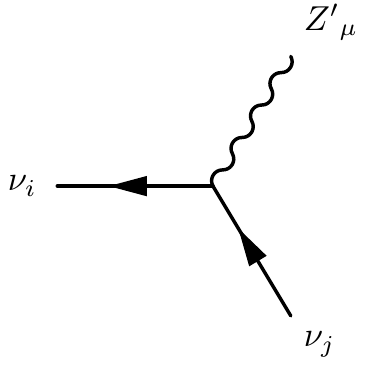}
\end{figure}. 
\begin{eqnarray*} 
 \frac{i}{2}\left(\left(\frac{e \, s_{\theta'}}{2 \, s_{W} c_{W}} + \frac{g_x'}{2} \, c_{\theta'}\right)\sum_{k=1}^{3} {\cal N}_{ik}{\cal N}^*_{jk} -  g_x \, c_{\theta'} \left(-\sum_{k=6}^{9} {\cal N}_{ik}{\cal N}^*_{jk}+\sum_{k=4}^{6} {\cal N}_{ik}{\cal N}^*_{jk}\right)\right)\!\gamma^\mu P_L \\
 -  \frac{i}{2}\left(\left(\frac{e \, s_{\theta'}}{2 \, s_{W} c_{W}} + \frac{g_x'}{2} \, c_{\theta'}\right)\sum_{k=1}^{3} {\cal N}_{ik}^*{\cal N}_{jk} -  g_x \, c_{\theta'} \left(-\sum_{k=6}^{9} {\cal N}^*_{ik}{\cal N}_{jk}+\sum_{k=4}^{6} {\cal N}^*_{ik}{\cal N}_{jk}\right)\right)\!\gamma^\mu P_R 
 \end{eqnarray*}

\noindent
where ${\cal N}$ is the neutrino mixing matrix as defined in Eq.~(\ref{VCKM2}). We note that $\nu_i$ for $i=1,2,3$ are identified as the light neutrinos and rest are heavy neutrinos. These neutrinos are Majorana fermions written in four-component notation.

\section*{Acknowledgments}
The authors would like to acknowledge support from the Department of Atomic Energy, Government of India, for the 
Regional Centre for Accelerator-based Particle Physics~(RECAPP). W.A. acknowledges support from the XII Plan Neutrino Project of the Department of Atomic Energy. T.S. acknowledges useful discussions with Nivedita Ghosh.
\bigskip

%

\end{document}